\documentclass[english, onecolumn, useAMS, usenatbib]{mn2e}
\usepackage{graphicx}
\usepackage{amsfonts}
\usepackage{subfigure}
\usepackage{color}
\usepackage{times}
\usepackage{amsmath}
\usepackage{babel}
\usepackage{csquotes}
\usepackage{amssymb}
\MakeOuterQuote{"}

\usepackage{ifpdf}
\ifpdf
\usepackage[hyperfootnotes=false,colorlinks=true]{hyperref}
\hypersetup{linkcolor=red, citecolor=blue}
\usepackage{epstopdf}
\fi

\def\lsim{\mathrel{\rlap{\lower4pt\hbox{\hskip1pt$\sim$}}
    \raise1pt\hbox{$<$}}}                
\def\gsim{\mathrel{\rlap{\lower4pt\hbox{\hskip1pt$\sim$}}
    \raise1pt\hbox{$>$}}}                

\begin{document}
\title[CMB polarization TE power spectrum estimation with ...]
{CMB polarization TE power spectrum estimation with non-circular beam}

\author[F. A. Ramamonjisoa et al.]
{Fidy A. Ramamonjisoa,$^{1}$ Subharthi Ray,$^{1}$ Sanjit Mitra,$^{2}$ and Tarun Souradeep$^{2}$\\
$^1$ Astrophysics and Cosmology Research Unit, School of
Mathematics, Statistics and Computer Science, University of KwaZulu-Natal, Private Bag X54001, \\
Durban 4000, South Africa\\
$^2$ Inter-University Centre for Astronomy and Astrophysics,
Post Bag 4, Ganeshkhind, Pune 411007, India\\
 }
\maketitle

\begin{abstract}
Precise measurements of the Cosmic Microwave Background (CMB) anisotropy have been one of the foremost concerns in modern cosmology as it provides valuable information on the cosmology of the universe. However, an accurate estimation of the CMB power spectrum faces many challenges as the CMB experiments sensitivity increases. Furthermore, for polarization experiments the precision of measurement is complicated by the fact that the polarisation signal is very faint compared to the measured total intensity, and could be impossible to detect in the presence of high level of systematics. One of the most important source of errors in CMB polarization  experiment is the beam asymmetry. For large data set the estimation of the CMB polarization power spectrum with standard optimal Maximum Likelihood (ML) is prohibitive for high resolution CMB experiments due to the enormous required computation time. In this paper, we present a semi-analytical framework using the pseudo-$C_{l}$ estimator to compute the power spectrum TE of the temperature anisotropy and the E-component of the polarization radiation field using non-circular beams. We adopt a model of beams obtained from a perturbative expansion of the beam around a circular (axisymmetric) one in harmonic space, and compute the resulting bias matrix which relates the true power spectrum with the observed one using an efficient algorithm for rapid computation. We show that for a multipole up to $l_{max}=500$,  the bias matrix can be computed in less than one second with a single CPU processor at 2.53 GHz. We find that the uncertainties induced by the beam asymmetry in the polarization power spectrum at the peak of the bias matrix for \textit{WMAP} and \textit{Planck} experiments can be as large as a few 10 to 20\%  (upper limit for \textit{Planck} LFI 30 GHz).
\end{abstract}


\section{Introduction}
\label{Sec: introduction}
Over the past decade, the data from the \textit{Wilkinson Microwave Anisotropy Probe} (\textit{WMAP}) \footnote{http://map.gsfc.nasa.gov/} \citep{Bennett:2003} programme has kept scientists busy analysing them, tuning the standard model of cosmology as to the finer details and testing various models of cosmology - a true decade of precision cosmology. With the launch of the \textit{Planck} \footnote{http://sci.esa.int/planck} satellite by the European Space Agency in 2009, and with data currently starting to pour in, it becomes more challenging to analyse the data, as \textit{Planck} probe for a much smaller angular resolution (larger multipoles). \textit{Planck} is scanning the CMB sky to multipoles of $\sim$ 3000, as compared to 1000 by the \textit{WMAP}. To make full use of the potential of the data that we receive it is necessary to accurately determine the observed data, eliminating the systematic effects. One of the primary objectives to probe the CMB sky to such high multipoles is to determine the polarized CMB signals and, in particular, the possibility of detection of the B-mode generated by tensor modes in the primordial perturbations of the density field which could indicate the signature of gravitational waves predicted by the cosmological inflation theory (\citealt{Kamionkowski:1997}; \citealt{Seljak:1997}). The CMB polarized signals (E-mode) produced by the quadrupolar anisotropy during the recombination epoch are a few order of magnitudes smaller than the total intensity of the anisotropy field (e.g., \citealt{Rees:1968}) implying that even an unsignificant systematic effects drastically bias the measurements of the CMB polarization. The asymmetric beam response of instruments, consequence of the off-axis placement of the detectors in telescopes, is one of the major systematic issue in CMB experiments as for small angular beam size, a highly asymmetric beam can strongly correlate with the sky signal distorting significantly the underlying true sky signal that we are unable to measure directly from observations. In CMB experiments the beam shape is measured using planets or bright sources observations combined with optical models (e.g., \citealt{Page:2003}; \citealt{Crill:2003}; \citealt{Masi:2006}; \citealt{Huffenberger:2010}). The actual beam pattern of the experiment can be complex (e.g., \textit{Archeops} \citealt{Macias:2007}) and special tools and techniques have been depicted to model the beam shapes (e.g., \citealt{Chiang:2002}; \citealt{Gorski:2005}; \citealt{Ashdown:2009}; \citealt{Huffenberger:2010}), but in general an elliptical Gaussian fit provides a good description of the main beam in many experiments (e.g., \textit{MAXIMA-1} \footnote{http://www.cfpa.berkeley.edu/group/cmb/} \citealt{Wu:2001}, \textit{Python V} \citealt{Souradeep:2001}, \textit{WMAP} \citealt{Page:2003}; \citealt{Jarosik:2007}; \citealt{Hill:2009}) and in particular, the ongoing \textit{Planck} survey (e.g., \citealt{Maffei:2010}; \citealt{Huffenberger:2010}; \citealt{Mitra:2011}; \citealt{PlanckIV:2013}; \citealt{PlanckVII:2013}). The far sidelobes can be modelled separately in spherical harmonics (e.g., \citealt{Wandelt:2001}), but we will neglect their contributions (see, e.g., \citealt{Burigana:2006}) as in this paper we mainly focus our attention on the computation of the coupling matrix between multipoles arising from the beam asymmetry. We note that sidelobe pickup systematic is $\sim$ 0.5 \% to 3.7\% of the total sky signal sensitivity for \textit{WMAP} \citep{Page:2003}. When the beam in CMB experiments is treated as circular (axisymmetric), it introduces systematic errors and biases the estimation of the power spectrum as the  imperfection of the instruments optics that triggers the asymmetry is unavoidable.\\
The use of the optimal ML is desirable as it provides an accurate estimation of the CMB angular power spectrum $C_{l}$. Different ML estimators have been implemented in data analysis, mainly for small data size (e.g., \citealt{Gorski:1994}; \citealt{Gorskietal.:1994}; \citealt{Tegmark:1997}; \citealt{Bond:1998}). These estimators can handle various systematics including correlated noise, non-uniform/cut-sky and asymmetric beams. The method consists to find the covariance that maximizes the likelihood function, defined as the integral over all possible values of the true temperature anisotropy $\Delta T(\hat q)$ ($\hat q$ denotes the pointing direction across the sky), which is statistically Gaussian distributed on the sky map. Nevertheless, the standard ML estimator requires intensive computation either in pixel space \citep{Bond:1998}, or Fourier space (\citealt{Gorski:1994}; \citealt{Gorskietal.:1994}). In fact, the evaluation  of the inverse of the covariance matrix of the likelihood that scales as $\sim O(N_{d}^{3})$ for a data size $N_{d}$ (see, e.g., \citealt{Borrill:1999}; \citealt{Bond:1999}), is computationally expensive, even impossible for large datasets such as \textit{WMAP} or \textit{Planck}. Various techniques have been developped to speed up the ML estimation, such as exploiting the scanning strategy symmetries (\citealt{Oh:1999}),  using hierarchical decomposition of the CMB map with  
varying degrees of resolution \citep{Dore:2001}, iterative multigrid method \citep{Pen:2003}. Specific methods such as estimate of the spectra on rings in the sky, can reduce the computational cost in special cases (\citealt{Challinor:2002}; \citealt{van:2002}; \citealt{Wandelt:2003b}). Other 'exact' power spectrum estimation methods have been proposed in the literature (e.g., \citealt{Knox:2001}; \citealt{Wandelt:2003a}; \citealt{Jewell:2004}).\\
Alternatively, the suboptimal pseudo-$C_{l}$ estimator \citep{Wandelt:2001a} provides a very convenient way for a fast computation ($\sim O(N_{d}^{\frac{3}{2}})$) of the angular power spectrum $C_{l}$ that we strive to measure. The pseudo-$C_{l}$ method is exploiting the fast spherical harmonic transform ($\sim N_{d}^{\frac{3}{2}}$) to estimate the angular power spectrum $C_{l}=\sum_{m}^{}a_{lm}^{T}a_{lm}^{E\ast}/(2l+1)$ (\citealt{Yu:1969}; \citealt{Peebles:1973}) from the data. This quadratic estimator is biased by the instrumental systematics that incorporate the beam asymmetry, partial sky/non-uniform sky coverage, noise, etc. Appropriate corrections of the systematic errors must be effectuated in order to guarantee the debiasing of the power spectrum estimator. Quadratic estimators can be assorted in the way  the spectra are corrected. We can illustrate the Spatially Inhomogenous Correlation Estimator (SpICE, \citealt{Szapudi:2001}) which computes the two-point correlation function $C(\hat q.\hat q')=\langle \Delta T(q) \Delta T(q')\rangle$ in pixel space in order to account for the non-uniformity of the sky coverage. \cite{Chon:2004} have proposed the extension of SpiCE to polarization. \cite{Wandelt:2001a} and \cite{Hivon:2002} have presented the Monte Carlo apodized spherical transform estimator (MASTER), which allows a fast computation of the angular power spectrum from the data before correcting the galactic cut in spherical harmonic space, by assuming circular beams. The method has been extended to the polarization (e.g., \citealt{Hansen:2003}; \citealt{Challinor:2005}; \citealt{Brown:2005}). \cite{Efstathiou:2004} and \cite{Efstathiou:2006} have suggested a hybrid algorithm that computes the power spectrum using ML for low-resolution maps at low multipoles, and pseudo-$C_{l}$ estimator for higher \textit{l} where it tends to be nearly optimal in the presence of dominant instrumental noise. This hybrid approach has been applied to the WMAP 3-yr data analysis \citep{Hinshaw:2007}. \cite{Hansen:2003} have employed the Gabor transforms to recover the power spectrum of the temperature and polarization on cut-sky, and \cite{Wandelt:2004} have exploited a global, exact method for power spectra recovery from CMB observations using Gibbs sampling \citep{Eriksen:2004}. The beam non-circularity (asymmetry) effects can be simulated into the covariance functions in approaches related to ML estimation (\citealt{Tegmark:1997}; \citealt{Bond:1998}), and can be included in Harmonic ring \citep{Challinor:2002} and ring-torus estimators \citep{Wandelt:2003b}. However, these estimators are computationally intensive and unfeasible for high-resolution maps and, the pseudo-$C_{l}$ method, sufficiently fast, is preferred for extracting the power spectrum at large multipoles \citep{Efstathiou:2004}.\\
The beam systematics have been investigated using semi-analytical methods based on the pseudo-$C_{l}$ estimator  (see, \citealt{Souradeep:2001}; \citealt{Fosalba:2002}; \citealt{Mitra:2004}; \citealt{Hinshaw:2007}; \citealt{Mitra:2009}; \citealt{Mitra:2011}) and full numerical integration (\citealt{Burigana:1998}; \citealt{Wu:2001}). The present work is focused on the optimization of the computation time of the bias matrix and, investigation of the effect of non-circular beams on the temperature and polarization (E-mode) power spectrum estimation with full sky coverage and noiseless limit, using the pseudo-$C_{l}$ method approach. We propose to extend the results of \cite{Fosalba:2002} by introducing the analytical tools developped in \cite{Mitra:2009} and, compute the bias matrix that encodes the power coupling at different multipoles that arises from the non-circularity of the beam patterns.\\
The paper is structured as follows. In Section \ref{Sec: formalism} we present the formalism which leads to the analytical expression of the multipole expansion of the temperature fluctuations and the E-mode polarization in harmonic space, that will allow the estimation of the TE power spectrum through the two-point correlation function using the pseudo-$C_{l}$ estimator. In Section \ref{Sec: Beam} we review the model of the beam harmonic transforms of the total intensity and the linear polarized radiation. We derive in Section \ref{Sec: Bias} the general expression of the bias matrix using a non-circular beam in the case of equal declination scan and for the particular case of non-rotating beam. We outline in Section \ref{Sec: Numeric} the numerical implementation of the bias matrix for non-rotating beam and evaluate the computation cost. In Section \ref{Sec: non-circular} we investigate the effect of the non-circularity of the beam and estimate the uncertainties on the power spectrum with respect to a circular Gaussian beam. We present in Section \ref{Sec: conclusion} the discussion and conclusion. We check the consistency of the bias matrix expression in the case of a circular (symmetric) beam in Appendix (\ref{appendix: checks}). We treat in Appendix (\ref{appendix: Integrals}) the evaluation of the integrals $\rm I_{1}$, $\rm I_{2}$, and $\rm I_{3}$ involved in the derivation of the bias matrix (Section \ref{Sec: Bias}) and in Appendix (\ref{appendix: Decomposition}) we give explicitly the decomposition of the general expression of the bias matrix of the TE power spectrum. 

\section{Formalism}
\label{Sec: formalism}
For the present formalism, we follow the approach developed in the paper of \cite{Mitra:2009} for the treatement of the total intensity of the radiation field and combine it with the semi-analytical treatment of the polarized CMB radiation developed in the paper of \cite{Fosalba:2002}. The idea consists to expand the total intensity of the anisotropy field and the polarized components of the CMB in harmonic space and cross-correlate the multipole coefficients of the expansion in order to obtain a simple theoretical estimate of the power spectrum. The ensemble average will be used to construct the pseudo-$C_{l}$ estimator. We expand the statistically Gaussian and isotropic temperature fluctuations $\Delta T(\hat q)$ over all sky directions $\hat{q}=(\theta, \phi),$  on the basis of the spherical harmonics as
\begin{eqnarray}
\label{eq: deltaT}
\Delta T(\hat q)=\sum_{lm}^{} a_{lm}^{T}Y_{lm}(\hat q)=\sum_{l=2}^{l_{max}}\sum_{m=-l}^{m=l} a_{lm}^{T}Y_{lm}(\hat q)
\end{eqnarray}
where $a_{lm}^{T}$ are the the coefficients of the temperature expansion in harmonic space. We apply the complex conjugate $Y^{\ast}_{l'm'}(\hat q)$ of the spherical harmonic function in Eq. (\ref{eq: deltaT}) and integrate over the solid angle of the sky $ \Omega_{\hat q} $ to derive
\begin{eqnarray}
 \int d\Omega_{\hat q} \Delta T(\hat q)Y^{\ast}_{l'm'}(\hat q)=\sum_{lm}^{}a_{lm}^{T}\int d\Omega_{\hat q}Y_{lm}(\hat q)Y^{\ast}_{l'm'}(\hat q),
\end{eqnarray}
and using the orthogonality and normalization relation (e.g., \citealt{Varshalovich:1988}) of the spherical harmonic function
\begin{eqnarray}
\int d\Omega_{\hat q}Y_{lm}(\hat q)Y^{\ast}_{l'm'}(\hat q)=\delta_{l l'}\delta_{m m'}
\end{eqnarray}
in which $\delta_{i j}$ denotes the Kronecker symbol, we obtain
\begin{eqnarray}
 a_{lm}^{T}= \int d\Omega_{\hat q} \Delta T(\hat q)Y^{\ast}_{lm}(\hat q)
\end{eqnarray}
where $d\Omega_{\hat q}=\rm sin\theta\it d\theta d\phi$. The above expression defines the multipoles $ a_{lm}^{T}$  as a function of the true temperature of the radiation field in spherical harmonic basis under an ideal systematic errors-free experiment assumption. In reality CMB experiments can only measure a disturbed temperature $\widetilde {\Delta T} (\hat q)$ triggered by instrument systematic effects. In this condition the coefficients of the harmonic transform of the observed temperature  $\widetilde {\Delta T} (\hat q)$  take the form
\begin{eqnarray}
\label{eq: harmo_coeff}
\tilde a_{lm}^{T}= \int d\Omega_{\hat q} \widetilde {\Delta T} (\hat q)Y^{\ast}_{lm}(\hat q).
\end{eqnarray}
As we focus our study on the effects of the beam non-circularity (asymmetry) in CMB surveys, we will neglect other systematics such as the cut-sky or non-uniformity of the sky due to the galactic mask applied in foreground removal processes.  In the case of full sky and noiseless limit the observed temperature $\widetilde {\Delta T} (\hat q)$ is the convolution of the true temperature $\Delta T(\hat q)$ with the beam as 
\begin{eqnarray}
\label{eq: obs_temp}
\widetilde {\Delta T} (\hat q)=\int d\Omega_{\hat q'} B(\hat q, \hat q')\Delta T (\hat q')
\end{eqnarray}
where $B(\hat q, \hat q')$ denotes the beam function. From Eq. (\ref{eq: harmo_coeff}) and Eq. (\ref{eq: obs_temp}) it follows that
\begin{eqnarray}
\label{eq: alm_expans}
\tilde a_{lm}^{T}= \int d\Omega_{\hat q}\int d\Omega_{\hat q'} B(\hat q, \hat q')\Delta T (\hat q')Y^{\ast}_{lm}(\hat q).
\end{eqnarray}
We expand $\Delta T (\hat q')$  in the spherical harmonic space as 
\begin{eqnarray}
\Delta T(\hat q')=\sum_{l'm'}^{} a_{l'm'}Y_{l'm'}(\hat q')
\end{eqnarray}
and plugg in Eq. (\ref{eq: alm_expans}) to derive the following form of the mutipole coefficients 
\begin{eqnarray}
\label{eq: almtildeT}
\tilde a_{lm}^{T}=\sum_{l'm'}^{} a_{l'm'}^{T}\int d\Omega_{\hat q}Y^{\ast}_{lm}(\hat q)\int d\Omega_{\hat q'} B(\hat q, \hat q')Y_{l'm'}(\hat q').
\end{eqnarray}
We can transform the above Eq. (\ref{eq: almtildeT}) by using the results of  the integration of the beam function as referred to Eq. (24) of \cite{Mitra:2009}
\begin{eqnarray}
\label{eq: Wigner}
\int d\Omega_{\hat q'} B(\hat q, \hat q')Y^{\ast}_{l'm'}(\hat q')&=&\sqrt{\frac{2l'+1}{4\pi}} \sum_{m''=-l'}^{l'}B_{l'}\beta_{l'm''}D_{m'm''}^{l'}(\hat q,\rho(\hat q)),\nonumber\\
\int d\Omega_{\hat q'} B(\hat q, \hat q')Y^{\ast}_{l'm'}(\hat q')&=& \sum_{m''=-l'}^{l'}b_{l'm''}^{T}D_{m'm''}^{l'}(\hat q,\rho(\hat q))
\end{eqnarray}
where
\begin{eqnarray}
\label{eq: Bl}
 B_{l}=\int_{-1}^{1} d(\hat q.\hat q')P_{l}(\hat q.\hat q')\left[ \frac{1}{2\pi}\int_{0}^{2\pi} d\phi\ B(\hat z, \hat q)\right] .
\end{eqnarray}
The beam distortion parameter defined as $\beta_{lm}={b_{lm}^{T}}/{b_{l0}}$ describes the deviation of the beam from circularity and
\begin{eqnarray}
 b_{lm}^{T}=\int d\Omega_{\hat q}\ Y^{\ast}_{lm}(\hat q)B(\hat z, \hat q)
\end{eqnarray}
denotes the beam harmonic transform of the total intensity of the underlying anisotropy field.
The terms $ D_{m'm''}^{l'}(\hat q,\rho(\hat q)) $ in Eq. (\ref{eq: Wigner}) are the Wigner-D functions given in terms of the Euler angles $(\theta, \phi, \rho)$.  The rotation angle $\rho(\hat q)$ describes the rotation of the beam along the pointing direction $\hat q$ whereas $D_{m'm''}^{l'}(\hat q,\rho(\hat q)) $ accounts for the rotation that carries the pointing direction $\hat q$ to the North Pole $\hat z$ axis \citep{Mitra:2004}. Eq. (\ref{eq: Bl}) includes the Legendre polynomials $ P_{l}(\hat q.\hat q') $ and the instrument beam response $ B(\hat z, \hat q) $ along the pointing direction $\hat z$. From Eq. (\ref{eq: almtildeT}) and Eq. (\ref{eq: Wigner}) we derive the general form of the temperature harmonic transform as a function of the intensity beam harmonic transform $b^{T}_{lm}$ as
\begin{eqnarray}
\label{eq: almTstar}
\tilde a_{lm}^{T\ast}&=&\sum_{l'm'}^{} a_{l'm'}^{T\ast}\sum_{m''=-l'}^{l'}b_{l'm''}^{T} \int d\Omega_{\hat q}Y_{lm}(\hat q)D_{m'm''}^{l'}(\hat q,\rho(\hat q))
\end{eqnarray}
which will be used to estimate the cross-power spectrum TE.\\
For the E-component of the polarization field we calculate the harmonic transform $\tilde a_{lm}^{E}$ using \cite{Fosalba:2002} approach by introducing the beam smoothed Stokes parameters $Q_{eff}$ and $U_{eff}$ on the spherical polar basis.
 The convolution of the beam with the sky can be expressed using Eq. (31) and Eq. (32) of \cite{Fosalba:2002} as
 \begin{eqnarray}
 \label{eq: Qeff}
 Q_{eff}&=& 2\sum_{lmM}[D_{mM}^{l}(\hat q,\rho(\hat q))]^{\ast}\ b_{lM}^{E\ast}\ a_{lm}^{E}\\
  \label{eq: Ueff}
 U_{eff}&=& 2\sum_{lmM}[D_{mM}^{l}(\hat q,\rho(\hat q))]^{\ast}\ b_{lM}^{E\ast}\ a_{lm}^{B}
 \end{eqnarray}
where $\lvert m\rvert \leq l$,  $\lvert M\rvert \leq l$. We adopt the E and B mode notation in our formula which  is connected to the gradient (G) and curl (C) components as $a_{lm}^{E}=-\sqrt{2}\ a_{lm}^{G}$ and $a_{lm}^{B}=-\sqrt{2}\ a_{lm}^{C}$.
In the case of  a symmetric (circular Gaussian) beam the Stokes parameters can be expanded in the spin-2 spherical harmonic basis $_{\ \mp 2}Y_{lm}(\hat q)$ as
\begin{eqnarray}
 (Q\pm iU)(\hat q)=\sum_{lm}(a_{lm}^{E}\mp i a_{lm}^{B})_{\ \mp 2}Y_{lm}(\hat q)
\end{eqnarray}
from which we derive the multipole coefficients of the polarization E-component given by
\begin{eqnarray}
\label{eq: almEsymmetric}
a_{lm}^{E}=\frac{1}{2}\int d\Omega_{\hat q}[(Q- iU)(\hat q)_{\ 2}Y^{\ast}_{lm}(\hat q)+(Q+ iU)(\hat q)_{\ -2}Y^{\ast}_{lm}(\hat q)].
\end{eqnarray}
We can obtain the sky multipoles of the non-circular beam $\tilde a_{lm}^{E}$ by plugging in Eq. (\ref{eq: almEsymmetric}) the effective smoothed Stokes parameters defined in Eq. (\ref{eq: Qeff}) and Eq. (\ref{eq: Ueff}) and after some algebra the final expression of the multipoles $\tilde a_{lm}^{E}$ reduces to 
\begin{eqnarray}
\label{eq: almEfinal_form}
\tilde a_{lm}^{E}&=&\int d\Omega_{\hat q}\sum_{l'm'}\sum_{M=-l'}^{l'}[D_{m'M}^{l'}(\hat q,\rho(\hat q))b_{l'M}^{E}]^{\ast}\ [a_{l'm'}^{E}(_{\ 2}Y^{\ast}_{lm}(\hat q)+ _{\ -2}Y^{\ast}_{lm}(\hat q))\nonumber\\
&-& ia_{l'm'}^{B}(_{\ 2}Y^{\ast}_{lm}(\hat q)- _{\ -2}Y^{\ast}_{lm}(\hat q))] .
\end{eqnarray}
From Eq. (\ref{eq: almTstar}) and Eq. (\ref{eq: almEfinal_form}) we can cross-correlate the temperature and the E-mode polarization in spherical harmonic space and get an estimate of the cross-power spectrum using the pseudo-$C_{l}$ estimator which will be defined later. In order to compute the two-point correlation function we need to evaluate the beam spherical harmonic transform of the total intensity of the field and the polarized one. In reality we can obtain this beam corrections from the CMB experiment but we can also simulate the beam response of the instrument. In the next Section \ref{Sec: Beam} we use the approach of \cite{Fosalba:2002} to describe the beam asymmetry model using the flat-sky approximation ($\theta\ll 1\rm\ rad$). 
\section{Beam spherical harmonic transform}
\label{Sec: Beam}
\begin{figure}
\begin{center}
\includegraphics [scale=0.6]{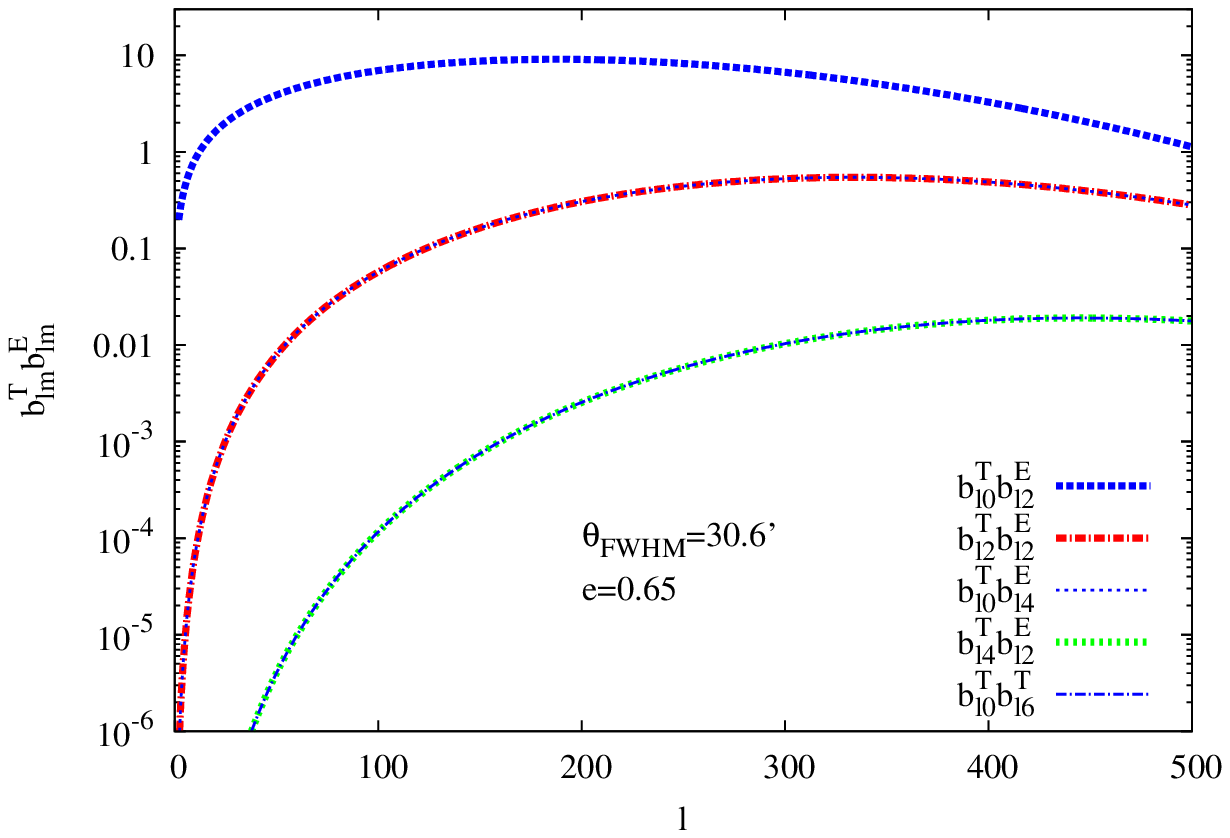}
\includegraphics[scale=0.6]{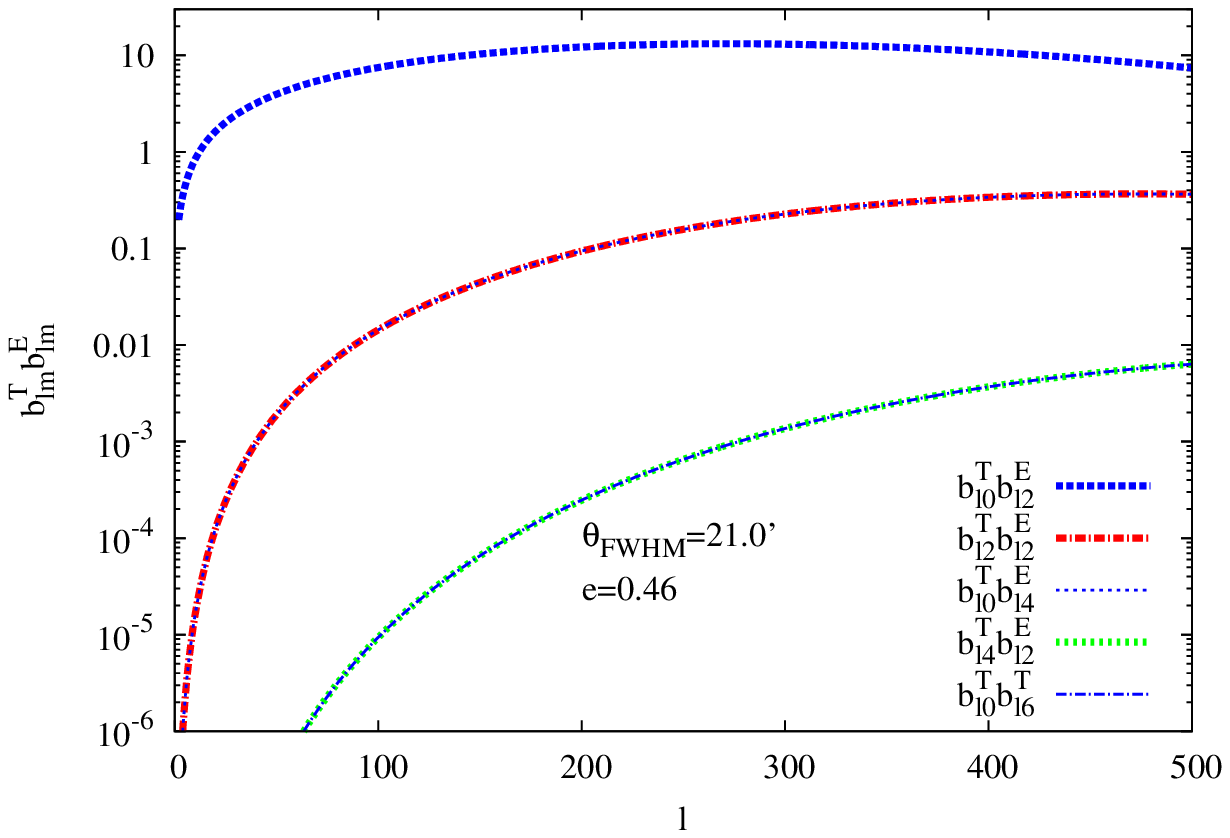}
\includegraphics[scale=0.6]{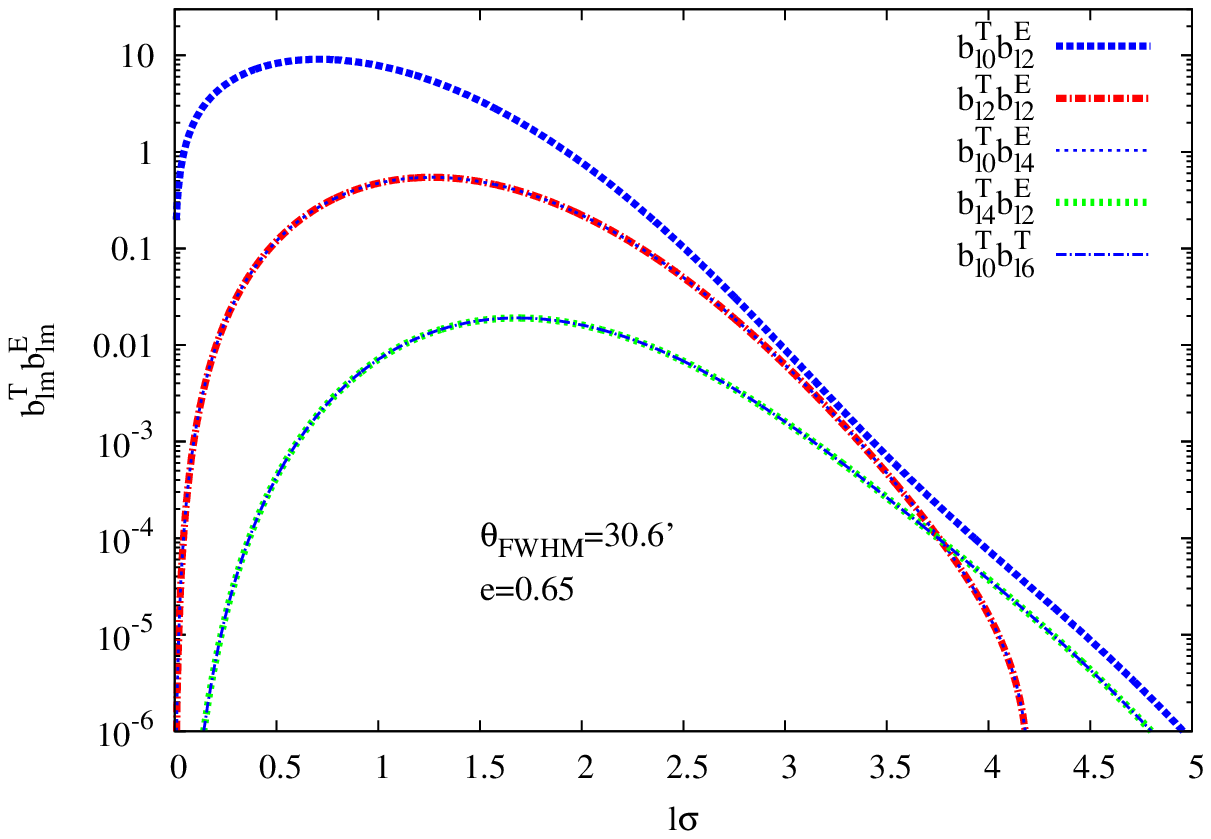}
\includegraphics[scale=0.6]{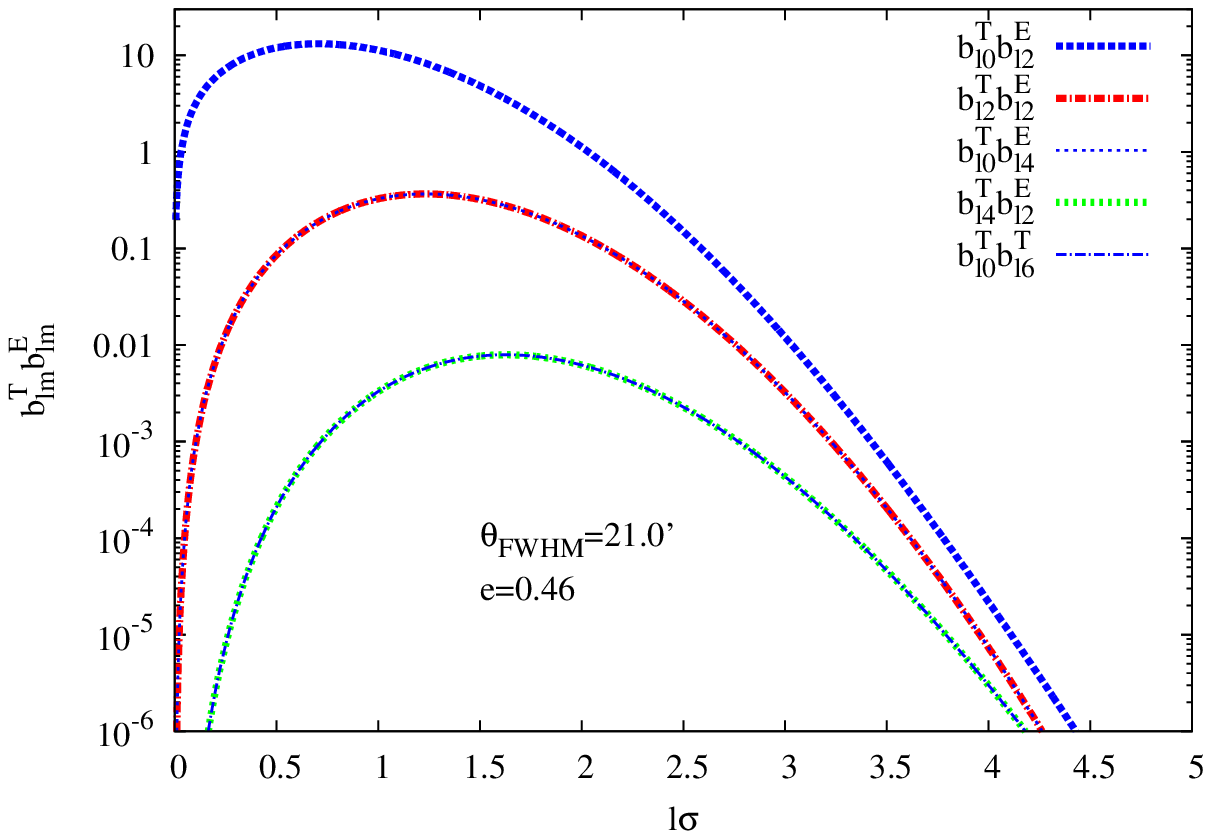}
\caption{Product of the temperature T and the E-polarized beam harmonic transforms. The top left panel shows the variation as a function of the multipole $l$ for a beam width $\theta_{\rm FWHM}=30.6'$ and eccentricity e=0.65 corresponding to the \textit{WMAP}- Q1 beam parameters model limited to $l_{max}=500$, and the bottom left panel illustrates the same product plotted against $l\sigma$. We show on the right panel the product of the beam transforms corresponding to the \textit{WMAP}- V beam with a beam width $\theta_{\rm FWHM}=21.0'$ and eccentricity e=0.46 plotted against the multipole $l$ (top right panel) and $l\sigma$ (bottom right panel). The first leading terms of the beam harmonic products (the blue thick dotted lines)  are connected to the circulary symmetric Gaussian window. The relation $b_{lm}^{E}=b_{lm-2}^{T}/2$ ($m$=2, 4, 6) implies that each of the second and third leading terms (non-circularity corrections) of the beam harmonic products incorporate two terms with the same amplitude which are illustrated with the overlapping dotted and dot-dashed lines. Note that the peak shifts to higher $l$ for higher order corrections and becomes important for $l\sigma \sim 1$.}
\label{Fig: blmTblmE}
\end{center}
\end{figure}

In this section we will use the results of  \cite{Fosalba:2002} which give the explicit forms of the beam spherical harmonic transforms of the intensity and the polarized beam. The approach is based on a perturbative expansion of an elliptical beam function $B(\theta, \phi)$ around 
a circular Gaussian beam in the flat-sky approximation which provides a good approximation for single-dish experiments.The beam window function is expanded in real and harmonic space from which a semi-analytic model of the beam harmonic transform can be obtained. Following \cite{Fosalba:2002}, the explicit form of the beam window function can be defined as
\begin{eqnarray}
B(\theta, \phi)=B_{0}\ \rm exp\left[ -\frac{\theta^{2}}{2\sigma_{\it b}^{2}} \it f(\phi)\right] 
\end{eqnarray}
in polar coordinates. The ellipticity parameter $\chi=1-(\sigma_{b}/\sigma_{a})^{2}$ of the window function is connected to
\begin{eqnarray}
f(\phi)=1-\chi\ \rm cos^{2}(\phi-\omega)
\end{eqnarray}
which describes the deviation of the beam from a circular (axisymmetric) one. $\sigma_{a}$ and $\sigma_{b}$ denote the beam widths along the major and minor axis. The expansion of the window function in spherical harmonic space reads
\begin{eqnarray}
B(\theta, \phi)=\sum_{l=2}^{l_{max}}\sum_{m=-l}^{l}b_{lm}Y_{lm}(\theta, \phi)
\end{eqnarray}
from which we obtain the beam harmonic transform $b_{lm}$
\begin{eqnarray} 
\label{eq: blmtransf}
b_{lm}=\int d\Omega B(\theta, \phi) Y_{lm}^{\ast}(\theta, \phi)
\end{eqnarray}
where $\Omega$ is the total solid angle over the sky. It is shown in \cite{Fosalba:2002} that Eq. (\ref{eq: blmtransf}) can be solved using a semi-analytical framework in the flat-sky approximation. Although a full numerical integration can also be performed (see, \citealt{Souradeep:2001}) to evaluate the integral involved in Eq. (\ref{eq: blmtransf} ) the rapidly converging semi-analytical approach is largely sufficient for our purpose. We will use the eccentricity $e=\sqrt{\chi}$ and the geometric mean beam width of the elliptic Gaussian window   
\begin{eqnarray}
\sigma (\rm in\ degrees) =\sqrt{\sigma_{a}\sigma_{b}}=\frac{\pi}{180}\frac{\theta_{FWHM}}{\sqrt{8\rm ln2}}
\end{eqnarray}
to describe accurately the beam geometry where $\theta_{\rm FWHM}$ denotes the full width at half maximum of the beam Gaussian profile.  The ellipticity of the beam will be defined by $\epsilon=\sigma_{a}/\sigma_{b}$ (not to be confounded  with the ellipticity parameter $\chi=1-1/\epsilon^{2}$). The harmonic expansion coefficients of the beam defined in Eq. (\ref{eq: blmtransf}) have symmetry properties which greatly simplify the numerical implementation. As mentioned in \cite{Fosalba:2002} only the even  $m$ modes have non-vanishing contribution in the beam transform as a consequence of the azimuthal symmetry of the term $\rm cos^{2}(\phi)$ which appears in the function $f(\phi)$. As a result of the property of the spherical harmonic complex conjugate $Y_{lm}^{\ast}(\theta, \phi)$ it follows that $b_{lm}^{\ast}=(-1)^{m}b_{l-m}$, and from the reality condition of the beam harmonic transform $b_{lm}^{\ast}=b_{lm}$.  This implies that $b_{lm}=b_{l-m}$ so that negative and positive modes have the same contribution. We will use the beam models of \cite{Fosalba:2002} that will allow us to compute numerically the bias matrix which describes the coupling between multipoles. For the temperature, the harmonic transform of the beam with second order in the parameter ellipticity $\chi$ is computed with the formula (A20) of \cite{Fosalba:2002} as follows
\begin{eqnarray}
\label{eq: beam_harmonics}
b^{T}_{l0}&=&\sqrt{\frac{2l+1}{4\pi}}e^{-\frac{l^{2}\sigma^{2}}{2}} \left[ 1-\frac{\chi}{4}l^{2}\sigma^{2}+\frac{\chi^{2}}{4}\left(  -l^{2}\sigma^{2}+ \frac{3}{16}l^{4}\sigma^{4}\right) \right],\nonumber\\
b^{T}_{l2}&=&\sqrt{\frac{2l+1}{4\pi}}\frac{\chi}{8}l^{2}\sigma^{2}e^{-\frac{l^{2}\sigma^{2}}{2}}\left[ 1+\chi\left( 1-\frac{1}{4}l^{2}\sigma^{2}\right) \right],\nonumber\\
b^{T}_{l4}&=&\sqrt{\frac{2l+1}{4\pi}}\frac{\chi^{2}}{128}l^{4}\sigma^{4}e^{-\frac{l^{2}\sigma^{2}}{2}},
\end{eqnarray}
and the linear polarized beam transforms with the same order expansion is obtained by using the simple relation $b^{E}_{l2}=b^{T}_{l0}/2$, $b^{E}_{l4}=b^{T}_{l2}/2$ and $b^{E}_{l6}=b^{T}_{l4}/2$.  We limit the perturbative expansion for both intensity and polarized beam to three terms which provides sufficient accuracy for experiments with mildly non-circular (asymmetric) beam. Under this prescriptions, the precision that can be achieved is $\sim$ 1\% till the multipole $l_{max}=5l_{peak}$ (for $\theta_{\rm FWHM}=10'$ and $\epsilon=1.3$) where $l_{peak}$ defined as $\sigma^{2}l^{2}_{peak}\simeq (1-\chi/4)$  is the multipole where the window function peaks (see, \citealt{Fosalba:2002}). As inferred from Eq. (\ref{eq: almTstar}) and Eq. (\ref{eq: almEfinal_form}), the correlation between T and E in harmonic space involves the product $b_{lm}^{T}b_{lm}^{E}$ which is illustrated in Fig. \ref{Fig: blmTblmE}. Clearly, the effect of higher order corrections ( $m$=2, 4 modes for T and $m$=4, 6 modes for E) due to the beam non-circularity (asymmetry) is significant for $l\sigma \sim 1$. Providing the analytical expansion of the beam total intensity and the E- polarized beams, we are able to compute the correlation between the sky harmonic coefficients defined  in Eq. (\ref{eq: almTstar}) and Eq. (\ref{eq: almEfinal_form}) and thereafter, the power spectrum of TE. We will use the pseudo-$C_{l}$ estimator whose advantage will be justified in the following Section \ref{Sec: Bias}.
\pagebreak
\section{The bias matrix}
\label{Sec: Bias}
Different methods have been proposed to estimate the power spectrum of CMB anisotropies. Among them, the optimal maximum likelihood estimator is the most commonly used but the huge computational cost makes it just unfeasible for high-resolution CMB experiments which probe small angular scales on the sky. Therefore, we adopt an alternative approach that will use the suboptimal  pseudo-$C_{l}$ estimator to compute the power spectrum. Whereas it is not an exact estimator as the maximum likelihood, it has the advantage  of being relatively  fast and can be used to process \textit{Planck}-like CMB large data sets in real time. The pseudo-$C_{l}$ estimator of the TE power spectrum is defined by
\begin{eqnarray}
\label{eq: pseudo_def}
 C^{TE}_{l}=\frac{1}{2l+1}\ \sum_{m=-l}^{l}a_{lm}^{T}a_{lm}^{E\ast}.
\end{eqnarray}
This suboptimal estimator is qualified as "pseudo" in the sense that it is biased. In order to obtain an accurate estimation of the CMB power spectra that corresponds to a real CMB experiment we need to take into account the systematic effects due to the beam asymmetry, the instrumental noise and  the non-uniformity and cut-sky \citep{Mitra:2004} as well as other systematics such as the telescope scanning and pointing errors, gain and calibration, power leakage between E and B-modes and the B-mode induced by the gravitational lensing of the CMB by large scale structure in the case of the B-mode polarization cross-spectra. We will investigate the effect of the beam asymmetry in the case of full sky coverage leaving apart for a future prospect the other systematic effects.\\
The expectation value of the pseudo-$C_{l}$ estimator is related to the true power spectrum $C_{l}^{TE}$ by
\begin{eqnarray}
\label{eq: C_TE_expectation_value}
\langle\tilde C^{TE}_{l}\rangle=\sum_{l'}^{}A_{ll'}^{TE}C_{l'}^{TE}
\end{eqnarray}
where $\langle\tilde C^{TE}_{l}\rangle$ denotes the ensemble average of the power spectrum over all realizations on the sky. The term $A_{ll'}^{TE}$ is the mutipoles coupling matrix which represents the bias (with respect to a cicular Gaussian profile) that will affect the estimation of the TE power spectrum  when the beam pattern is asymmetric.  From Eq. (\ref{eq: pseudo_def})  we can write the expectation value of the cross-power spectrum TE as 
\begin{eqnarray}
\label{eq: Cl_TE_expect}
\langle \tilde C^{TE}_{l}\rangle&=&\frac{1}{2l+1}\ \sum_{m=-l}^{l}\langle \tilde a_{lm}^{T}\tilde a_{lm}^{E\ast}\rangle\nonumber\\
&=&\frac{1}{2l+1}\ \sum_{m=-l}^{l}\langle \tilde a_{lm}^{T\ast}\tilde a_{lm}^{E}\rangle.
\end{eqnarray}
We substitute Eq. (\ref{eq: almTstar}) and Eq. (\ref{eq: almEfinal_form}) into Eq. (\ref{eq: Cl_TE_expect}) to obtain the ensemble average of the power spectrum which reads
\begin{eqnarray}
\langle \tilde C^{TE}_{l}\rangle &=&\frac{1}{2l+1}\sum_{m=-l}^{l}\sum_{l'm'l'_{1}m'_{1}}^{}\langle a_{l'm'}^{T\ast}a_{l'_{1}m'_{1}}^{E}\rangle \sum_{m''=-l'}^{l'}b^{T}_{l'm''}\int d\Omega_{\hat q} Y_{lm}(\hat q)D_{m'm''}^{l'}(\hat q,\rho(\hat q))\nonumber\\
&\times &\sum_{M=-l'_{1}}^{l'_{1}}b^{E}_{l'_{1}M}\int d\Omega_{\hat q}[D_{m'_{1}M}^{l'_{1}}(\hat q,\rho(\hat q))]^{\ast}(_{\ 2}Y^{\ast}_{lm}(\hat q)+\ _{-2}Y^{\ast}_{lm}(\hat q)).
\end{eqnarray}
Then we use the statistical isotropy of the CMB anisotropy
\begin{eqnarray}
\langle a_{l'm'}^{T\ast}\ a_{l'_{1}m'_{1}}^{E}\rangle&=&C^{TE}_{l'}\ \delta_{l'l'_{1}}\delta_{m'm'_{1}},\\
\langle a_{l'm'}^{T\ast}\ a_{l'_{1}m'_{1}}^{B}\rangle&=&C^{TB}_{l'}\ \delta_{l'l'_{1}}\delta_{m'm'_{1}}=0
\end{eqnarray}
to derive the following expression
\begin{eqnarray}
\label{eq: Cl_TE_result}
\langle \tilde C^{TE}_{l}\rangle &=&\frac{1}{2l+1}\sum_{m=-l}^{l}\sum_{l'm'}^{}C^{TE}_{l'} \sum_{m''=-l'}^{l'}b^{T}_{l'm''}\int d\Omega_{\hat q} Y_{lm}(\hat q)D_{m'm''}^{l'}(\hat q,\rho(\hat q))\nonumber\\
&\times &\sum_{M=-l'}^{l'}b^{E}_{l'M}\left[ \int d\Omega_{\hat q}(_{\ 2}Y_{lm}(\hat q)D_{m'M}^{l'}(\hat q,\rho(\hat q))+\ _{-2}Y_{lm}(\hat q)D_{m'M}^{l'}(\hat q,\rho(\hat q))\right] ^{\ast}\nonumber\\
 &=&\frac{1}{2l+1}\sum_{m=-l}^{l}\sum_{l'm'}^{}C^{TE}_{l'} \sum_{m''=-l'}^{l'}b^{T}_{l'm''}\rm I_{1} \sum_{M=-l'}^{l'}b^{E}_{l'M}[\rm I_{2}+\rm I_{3}]^{\ast}
\end{eqnarray}
where we define the integrals $\rm I_{1}$, $\rm I_{2}$, and $\rm I_{3}$ as
\begin{eqnarray}
\rm I_{1}&=&\int d\Omega_{\hat q}\ Y_{lm}(\hat q)D_{m'm''}^{l'}(\hat q,\rho(\hat q)),\\
\rm I_{2}&=&\int d\Omega_{\hat q}\ _{\ 2}Y_{lm}(\hat q)D_{m'M}^{l'}(\hat q,\rho(\hat q)),\\
\rm I_{3}&=&\int d\Omega_{\hat q}\ _{\ -2}Y_{lm}(\hat q)D_{m'M}^{l'}(\hat q,\rho(\hat q)).
\end{eqnarray}
The above integrals depend on the scanning strategy of the CMB survey through the function $\rho(\hat q)$ which defines the rotation of the beam along the pointing direction of the telescope. We will show that under some assumptions on  this scan pattern the integrals  $\rm I_{1}$, $\rm I_{2}$ and $\rm I_{3}$ can be solved analytically. We report in Appendix (\ref{appendix: Integrals}) the derivations of the integrals $\rm I_{1}$, $\rm I_{2}$, and $\rm I_{3}$. By replacing the integrals $\rm I_{1}$, $\rm I_{2}$ and $\rm I_{3}$ into Eq. (\ref{eq: Cl_TE_result}) we get the following expression of the expectation value of the power spectrum
\begin{eqnarray}
\langle \tilde C^{TE}_{l}\rangle &=&\frac{1}{2l+1}\sum_{m=-l}^{l}\sum_{l'm'}^{}C^{TE}_{l'} \sum_{m''=-l'}^{l'}b^{T}_{l'm''} (-1)^{m}\sqrt{\frac{2l+1}{4\pi}}\sum_{L=\vert l-l'\vert}^{l+l'}C_{l-m l' m'}^{L(-m+m')}C_{l 0 l' m''}^{L m''}\nonumber\\
&\times & \int d\Omega_{\hat q} D_{(-m+m')m''}^{L}(\hat q,\rho(\hat q))\sum_{M=-l'}^{l'}b^{E}_{l'M} (-1)^{m}\sqrt{\frac{2l+1}{4\pi}} \sum_{L'=\vert l-l'\vert}^{l+l'}C_{l-m l' m'}^{L'(-m+m')}\nonumber \\                         
&\times & \left[  C_{l2 l' M}^{L'(2+M)}  \int d\Omega_{\hat q}D_{(-m+m')(2+M)}^{L'}(\hat q,\rho(\hat q)) C_{l-2 l' M}^{L'(-2+M)}  \int d\Omega_{\hat q}D_{(-m+m')(-2+M)}^{L'}(\hat q,\rho(\hat q))\right]   ^{\ast}
\end{eqnarray}
which can be simplified in the form
\begin{eqnarray}
\label{eq: C_TEaverage}
\langle \tilde C^{TE}_{l}\rangle &=&\frac{1}{4\pi}\sum_{m=-l}^{l}\sum_{l'm'}^{}C^{TE}_{l'} \sum_{m''=-l'}^{l'}b^{T}_{l'm''} \sum_{L=\vert l-l'\vert}^{l+l'}C_{l-m l' m'}^{L(-m+m')}C_{l 0 l' m''}^{L m''}\nonumber\\
&\times & \int d\Omega_{\hat q} D_{(-m+m')m''}^{L}(\hat q,\rho(\hat q))\sum_{M=-l'}^{l'}b^{E}_{l'M}  \sum_{L'=\vert l-l'\vert}^{l+l'}C_{l-m l' m'}^{L'(-m+m')}\nonumber \\                         
&\times & \left[  C_{l2 l' M}^{L'(2+M)}  \int d\Omega_{\hat q}D_{(-m+m')(2+M)}^{L'}(\hat q,\rho(\hat q)) C_{l-2 l' M}^{L'(-2+M)}  \int d\Omega_{\hat q}D_{(-m+m')(-2+M)}^{L'}(\hat q,\rho(\hat q))\right]   ^{\ast}.
\end{eqnarray}
In the following we propose to compute the integral
\begin{eqnarray}
\chi_{mm'}^{l}[\rho(\hat q)]=\int d\Omega_{\hat q}D_{mm'}^{l}(\hat q, \rho(\hat q))
\end{eqnarray}
involved in Eq. (\ref{eq: C_TEaverage}) which  depends on the scanning strategy of the CMB experiment. We can evaluate $\chi_{mm'}^{l}[\rho(\hat q)]$ using \cite{Mitra:2009} approach. We consider a beam rotation $\rho(\hat q)$ which can be decomposed into declination and right ascension parts $\rho(\hat q)=\Theta(\theta)+\Phi(\phi)$ as it provides a good approximation of real scan strategies. Using Eq. (1), Section 4.3 of \cite{Varshalovich:1988}
\begin{eqnarray}
D_{m m'}^{l}(\hat q,\rho(\hat q))=e^{-im\phi}d_{m m'}^{l}(\theta)e^{-im'\rho},
\end{eqnarray}
where $d_{m m'}^{l}(\theta)$ denotes the Wigner-d function.
$\chi_{mm'}^{l}[\rho(\hat q)]$ takes the form
\begin{eqnarray}
\label{eq: khi_rho}
\chi_{mm'}^{l}[\rho(\hat q)]=\int_{0}^{2\pi}d\phi e^{-im'\Phi(\phi)} \int_{0}^{\pi}d\theta \ \rm sin\theta\ \it d_{mm'}^{l}(\theta)e^{-im'\Theta(\theta)}
\end{eqnarray}
which contributes significantly only for constrained values of $m'$.\\
For an equal declination scan $\rho(\hat q)=\rho(\theta)$, and the integral $\chi_{mm'}^{l}[\rho(\hat q)]$ reduces to
\begin{eqnarray}
\chi_{mm'}^{l}[\rho(\theta)]&=&2\pi \delta_{m0} \int_{0}^{\pi}d\theta \ \rm sin\theta\ \it d_{mm'}^{l}(\theta)e^{-im'\rho(\theta)}\nonumber\\
\label{eq: Khi}
&=&\chi_{0m'}^{l}[\rho(\theta)]
\end{eqnarray}
where the Wigner-d function $d_{0m'}^{l}(\theta)$ (since the only non-vanishing terms are obtained for $m=0$) can be evaluated using Eq. (E15) of \cite{Mitra:2009} as
\begin{eqnarray}
d_{0m'}^{l}(\theta)=i^{m'}\sum_{N=-l}^{l}\left[  (-1)^{N}d_{0N}^{l}(\frac{\pi}{2})e^{iN\theta}d_{Nm'}^{l}(\frac{\pi}{2})\right] .
\end{eqnarray}
Then 
\begin{eqnarray}
\chi_{0m'}^{l}[\rho(\theta)]&=& 2\pi \sum_{N=-l}^{l}d_{0N}^{l}(\frac{\pi}{2})d_{Nm'}^{l}(\frac{\pi}{2}) i^{m'}(-1)^{N}\int_{0}^{\pi}d\theta \ \rm sin\theta\ \it e^{iN\theta}e^{-im'\rho(\theta)}\nonumber\\
\label{eq: Gamma_mpn }
&=&2\pi \sum_{N=-l}^{l}d_{0N}^{l}(\frac{\pi}{2})d_{Nm'}^{l}(\frac{\pi}{2})\Gamma_{m'N}[\rho(\theta)]
\end{eqnarray}
where
\begin{eqnarray}
\label{eq: Gamma_m'N}
\Gamma_{m'N}[\rho(\theta)]=i^{m'}(-1)^{N}\int_{0}^{\pi}d\theta \ \rm sin\theta\ \it e^{iN\theta}e^{-im'\rho(\theta)}.
\end{eqnarray}
\cite{Mitra:2009} show that only the real parts of $\Gamma_{m'N}[\rho(\theta)]$ contribute in most of the cases where the general shape of the beam and the scan strategies exhibit trivial symmetries. For non-rotating beam $\rho(\hat q)=0$, and the real part of the $\Gamma$ coefficients reduces to Eq. (38) of \cite{Mitra:2009} expressed as follows
\begin{eqnarray}
\label{eq:deff}
\Re\left[\Gamma_{m'N}[\rho(\hat q)=0]\right] \ = \ f_{m'N} \ = \
\left\{
\begin{array}{cl} (-1)^{(m'\pm 1)/2} \, \pi/2        & \mbox{if
$m'$=odd and $N=\pm 1$}\\ (-1)^{m'/2}
\, 2/(1-N^2) & \mbox{if both $m',N=0$ or even}\\
0               & \mbox{otherwise.} \end{array} \right.
\end{eqnarray}
We apply the property obtained in Eq. (\ref{eq: Khi}) and the relation given in Eq. (\ref{eq: Gamma_mpn }) to derive the following quantities assuming that the beam is non-rotating.
\begin{eqnarray}
\int d\Omega_{\hat q}D_{(-m+m')m''}^{L}(\hat q, \rho(\hat q))&=&2\pi \sum_{M=-L}^{L}d_{0M}^{L}(\frac{\pi}{2})d_{Mm''}^{L}(\frac{\pi}{2}) f_{m''M}\\
\int d\Omega_{\hat q}D_{(-m+m')(2+M)}^{L'}(\hat q, \rho(\hat q))&=&2\pi \sum_{N=-L'}^{L'}d_{0N}^{L'}(\frac{\pi}{2})d_{N(2+M)}^{L'}(\frac{\pi}{2}) f_{(2+M)N}\\
\int d\Omega_{\hat q}D_{(-m+m')(-2+M)}^{L'}(\hat q, \rho(\hat q))&=&2\pi \sum_{N=-L'}^{L'}d_{0N}^{L'}(\frac{\pi}{2})d_{N(-2+M)}^{L'}(\frac{\pi}{2}) f_{(-2+M)N}.
\end{eqnarray}
Eq. (\ref{eq: Khi}) implies that the only non-vanishing terms from the above integrals are obtained for $m'=m$. After some algebra the general expression of the bias matrix for an equal declination scan strategy can be derived from Eq. (\ref{eq: C_TEaverage}) in the form
\begin{eqnarray}
\label{eq: All'gamma}
A_{ll'}^{TE}&=&\pi\sum_{m=-l}^{l} \sum_{m''=-l'}^{l'}b^{T}_{l'm''} \sum_{L=\vert l-l'\vert}^{l+l'}C_{l-m l' m}^{L 0}C_{l 0 l' m''}^{L m''} \sum_{M=-L}^{L}d_{0M}^{L}(\frac{\pi}{2})d_{Mm''}^{L}(\frac{\pi}{2}) \Gamma_{m''M}[\rho(\theta)]\sum_{M'=-l'}^{l'}b^{E}_{l'M'} \sum_{L'=\vert l-l'\vert}^{l+l'}C_{l-m l' m}^{L' 0}\nonumber\\
&\times &\left[  C_{l2 l' M'}^{L'(2+M')}\sum_{N=-L'}^{L'} d_{0N}^{L'}(\frac{\pi}{2}) d_{N(2+M')}^{L'}(\frac{\pi}{2}) \Gamma_{(2+M')N}[\rho(\theta)]\right. \nonumber\\
 &+& \left. C_{l-2 l' M'}^{L'(-2+M')}\sum_{N=-L'}^{L'}  d_{0N}^{L'}(\frac{\pi}{2})d_{N(-2+M')}^{L'}(\frac{\pi}{2}) \Gamma_{(-2+M')N}[\rho(\theta)]\right]
\end{eqnarray}
and for the particular case of a non-rotating beam ($\Gamma_{m'N}[\rho(\hat q)=0]\ = \ f_{m'N} $) the bias matrix takes the form
\begin{eqnarray}
\label{eq: Allprime_fmprimeN}
A_{ll'}^{TE}&=&\pi\sum_{m=-l}^{l} \sum_{m''=-l'}^{l'}b^{T}_{l'm''} \sum_{L=\vert l-l'\vert}^{l+l'}C_{l-m l' m}^{L 0}C_{l 0 l' m''}^{L m''} \sum_{M=-L}^{L}d_{0M}^{L}(\frac{\pi}{2})d_{Mm''}^{L}(\frac{\pi}{2})f_{m''M}\sum_{M'=-l'}^{l'}b^{E}_{l'M'} \sum_{L'=\vert l-l'\vert}^{l+l'}C_{l-m l' m}^{L' 0}\nonumber\\                      
&\times & \left[ C_{l2 l' M'}^{L'(2+M')}\sum_{N=-L'}^{L'} d_{0N}^{L'}(\frac{\pi}{2}) d_{N(2+M')}^{L'}(\frac{\pi}{2}) f_{(2+M')N}\right.\nonumber\\
&+&\left. C_{l-2 l' M'}^{L'(-2+M')}\sum_{N=-L'}^{L'}  d_{0N}^{L'}(\frac{\pi}{2})d_{N(-2+M')}^{L'}(\frac{\pi}{2}) f_{(-2+M')N}\right] 
\end{eqnarray}
where $m''=0, \pm2, \pm4 $ and $M'=\pm2, \pm4, \pm6 $ are the modes corresponding to the total intensity beam and the polarized beam transforms with second order in ellipticity. Eq. (\ref{eq: Allprime_fmprimeN}) constitutes one of the main results of this paper. It provides the most general form of the bias matrix for a non-circular (asymmetric)  beam in the case of a full sky coverage. We can see from Eq. (\ref{eq: All'gamma}) and Eq. (\ref{eq: Allprime_fmprimeN}) that the computational cost of the bias matrix for an equal declination scan is equivalent to that of the non-rotating beam as we need only to precompute the coefficients $\Gamma_{m'N}[\rho(\theta)]$ and $f_{m'N}$ for the corresponding scan strategies. As a result of the constraint on $m$ in Eq. (\ref{eq: Khi})  and as discussed in the paper of \cite{Mitra:2009}, we can expect that the bias matrix computation for real scan strategies is computationally equivalent to the bias computation for non-rotating beam. 
 \section{Numerical implementation}
 \label{Sec: Numeric}
 The bias matrix defined in Eq. (\ref{eq: Allprime_fmprimeN}) contains implicit information about the modes coupling between multipoles attributed to the beam asymmetry. A detail study of the effect of the non-circularity of the beam in the power spectrum estimation requires numerous and repeated computations of the bias matrix with different beam parameters (beam width and eccentricity) at each step of the computation. However, the numerical evaluation of the algebraic expression in Eq. (\ref{eq: Allprime_fmprimeN}) is a computational challenge. A naive numerical implementation of the formula Eq. (\ref{eq: Allprime_fmprimeN}) would scale as $O(l_{max}^{6})$ which becomes quickly prohibitive at large multipoles (smaller angular resolution). In this section we will tackle this issue and estimate the computational cost of the evaluation of the bias matrix. We first decompose the summations involved in Eq. (\ref{eq: Allprime_fmprimeN}) in order to split the bias matrix into several ones in which, separately appears the beam harmonic transforms product $b_{lm}^{T}b_{lm}^{E}$ introduced in Section \ref{Sec: Beam}. Therefore, Eq. (\ref{eq: Allprime_fmprimeN}) can be written as follows
 \begin{eqnarray}
 A_{ll'}^{TE}=A_{ll'}^{TE}(term\ 1)+ A_{ll'}^{TE}(term\ 2)+ A_{ll'}^{TE}(term\ 3)+ A_{ll'}^{TE}(term\ 4)+ A_{ll'}^{TE}(term\ 5)
 \end{eqnarray}
 where each term of the bias matrix derived from Eq. (\ref{eq: Allprime_fmprimeN}) is given explicitly in Appendix (\ref{appendix: Decomposition}). The first term  $A_{ll'}^{TE}(term\ 1)$ is the bias corresponding to the leading term $b_{l0}^{T}b_{l2}^{E}$ of the harmonics product. We will show in Appendix (\ref{appendix: checks}) that this term reduces to the well known window function for a circular symmetric beam which reads
 \begin{eqnarray}
  A_{ll'}^{TE}(term\ 1)=e^{-l^{2}\sigma^{2}}\delta_{ll'}.
 \end{eqnarray}
 Hereafter, we will introduce the symmetry properties of the Wigner-d functions that considerably allow us to  reduce the number of operations involved in the computation of the summations of kind $\sum_{N=-l}^{l}$. We define the quantities
 \begin{eqnarray}
 \label{eq: dL'm'}
 d(L', m')&=&\sum_{N=-L'}^{L'}d(L', N, m')
 \end{eqnarray}
 where 
 \begin{eqnarray}
 d(L', N, m')&=& d_{0N}^{L'}(\frac{\pi}{2})d_{Nm'}^{L'}(\frac{\pi}{2}) f_{m'N}
 \end{eqnarray}
 for $m'=\pm 2,\ \pm 4,\ \pm 6$. We notice that the non-rotating beam scanning strategy described by the function $f_{m'N}$ involved in Eq. (\ref{eq:deff}) is of even parity with respect to $N$, i.e., $f_{m'-N}=f_{m'N}$.  From the following symmetry relations of the Wigner-d function (Eq. (1), Section 4.4 of \citealt{Varshalovich:1988})
\begin{eqnarray}
 d_{0-N}^{L'}(\frac{\pi}{2})=(-1)^{N}d_{0N}^{L'}(\frac{\pi}{2})=(-1)^{L'}d_{0N}^{L'}(\frac{\pi}{2}),
 \end{eqnarray}
 \begin{eqnarray}
 d_{-Nm'}^{L'}(\frac{\pi}{2})=(-1)^{L'-m'}d_{Nm'}^{L'}(\frac{\pi}{2}),
 \end{eqnarray}
 we can write that
 \begin{eqnarray}
 d_{0-N}^{L'}(\frac{\pi}{2})d_{-Nm'}^{L'}(\frac{\pi}{2}) f_{m'-N}&=&(-1)^{m'}d_{0N}^{L'}(\frac{\pi}{2})d_{Nm'}^{L'}(\frac{\pi}{2}) f_{m'N}
 \end{eqnarray}
 which translates to 
 \begin{eqnarray}
 d(L', -N, m') =(-1)^{m'}d(L', N, m'),
 \end{eqnarray}
 and finally by plugging in Eq. (\ref{eq: dL'm'}), the above relation gives
 \begin{eqnarray}
 \label{eq: dL'm'sym}
 d(L', m')=d_{00}^{L'}(\frac{\pi}{2})d_{0m'}^{L'}(\frac{\pi}{2}) f_{m'0}+ \sum_{N=1}^{L'}((-1)^{m'}+ 1)d_{0N}^{L'}(\frac{\pi}{2})d_{Nm'}^{L'}(\frac{\pi}{2}) f_{m'N}
 \end{eqnarray}
 which is valid for the different values of $m'$. We can see from Eq. (\ref{eq: dL'm'sym}) that instead of $2L'$ additions, only $L'$ operations are necessary for the computation of each summation of the form defined by Eq. (\ref{eq: dL'm'}). In this way, we will gain a computational improvement by a factor of two. Analogously, the evaluation of the terms $A_{ll'}^{TE}(term\ 2),\ A_{ll'}^{TE}(term\ 3),\ A_{ll'}^{TE}(term\ 4)$ and $A_{ll'}^{TE}(term\ 5)$ can be carried out by following the same algebra formalism exposed in Appendix (\ref{appendix: checks}) where we treat a detailed derivation of the term $A_{ll'}^{TE}(term\ 1)$. Furthermore, we can reduce the number of addition operations by including the Clebsch-Gordan coefficients symmetry properties. We notice that each term of the bias matrix contains the summation $\sum_{m=-l}^{l}C_{l-m l' m}^{L 0}$ which can be written using Eq. (11), Section 8.4.3 of \cite{Varshalovich:1988} as
 \begin{eqnarray}
 \label{eq: cll'L}
 c(l, l', L)&=&\sum_{m=-l}^{l}C_{l-m l' m}^{L 0}\\
 &=&\sum_{m=-min(l,\ l')}^{min(l,\ l')}C_{l-m l' m}^{L 0}\\
 \label{eq: cll'Lsym}
& =&C_{l0 l' 0}^{L 0} + ((-1)^{L}+ 1)\sum_{m=1}^{min(l,\ l')}(-1)^{m}C_{l-m l' m}^{L 0},
 \end{eqnarray}
 which again reduces by a factor of two the operations needed for the summations. Putting all together and introducing the new notations introduced in  Eq. (\ref{eq: dL'm'}) and Eq. (\ref{eq: cll'L}), we can resume the expression of the different terms of the bias matrix as follows
 \begin{eqnarray}
 A_{ll'}^{TE}(term\ 1)=e^{-l^{2}\sigma^{2}}\ \delta_{ll'},
 \end{eqnarray}
\begin{eqnarray}
 A_{ll'}^{TE}(term\ 2)=\frac{4\pi}{2l+1}b_{l'2}^{T}b_{l'2}^{E}\sum_{L=\vert l-l'\vert}^{l+l'}C_{l0 l' 2}^{L 2}\ \left( (-1)^{L}d(L,-2)+ d(L, 2)\right) \ c(l, l', L),
\end{eqnarray}
 \begin{eqnarray}
 A_{ll'}^{TE}(term\ 3)=\frac{2\pi}{2l+1}b_{l'0}^{T}b_{l'4}^{E}\sum_{L=\vert l-l'\vert}^{l+l'} \left[C_{l-2 l' 4}^{L 2}\  \left( (-1)^{L}d(L,-2)+ d(L, 2)\right) +  C_{l2 l' 4}^{L 6}\  \left( (-1)^{L}d(L,-6)+ d(L, 6)\right) \right]c(l, l', L), 
\end{eqnarray}
 \begin{eqnarray}
 A_{ll'}^{TE}(term\ 4)=\frac{4\pi}{2l+1}b_{l'4}^{T}b_{l'2}^{E}\sum_{L=\vert l-l'\vert}^{l+l'}\ C_{l0 l' 4}^{L 4}\  \left( (-1)^{L}d(L,-4)+ d(L, 4)\right) c(l, l', L), 
\end{eqnarray}
\begin{eqnarray}
 A_{ll'}^{TE}(term\ 5)=\frac{2\pi}{2l+1}b_{l'0}^{T}b_{l'6}^{E}\sum_{L=\vert l-l'\vert}^{l+l'} \left[C_{l-2 l' 6}^{L 4}\  \left( (-1)^{L}d(L,-4)+ d(L, 4)\right) +  C_{l2 l' 6}^{L 8}\  \left( (-1)^{L}d(L,-8)+ d(L, 8)\right) \right]c(l, l', L). 
\end{eqnarray}  
In order to compute efficiently the bias matrix, we need to simplify as much as possible the above formula. We further proceed with the algorithm implementation by introducing the new quantities
\begin{eqnarray}
\label{eq: d2468L}
dm'(L)=(-1)^{L}d(L,-m')+ d(L, m')
\end{eqnarray}
which will greatly ease the computation as they appear several times in the bias matrix expressions ($m'=2,\ 4,\ 6,\ 8$). As the above quantities can be precomputed, we can expect a fast computation of the bias matrix in real time. At this step the summation of the bias matrix terms reduces to the following relation
\begin{eqnarray}
\label{eq: new_All'}
 A_{ll'}^{TE}&=&e^{-l^{2}\sigma^{2}}\ \delta_{ll'}\nonumber\\
 &+& \frac{4\pi}{2l+1}\sum_{L=\vert l-l'\vert}^{l+l'} \left[b_{l'2}^{T}b_{l'2}^{E}\ C_{l0 l' 2}^{L 2}d2(L)+ \frac{1}{2}b_{l'0}^{T}b_{l'4}^{E}\left( C_{l-2 l' 4}^{L 2}d2(L)+ C_{l2 l' 4}^{L 6}d6(L)\right)\right.\nonumber\\
 &+& \left. b_{l'4}^{T}b_{l'2}^{E}\ C_{l0 l' 4}^{L 4}d4(L)+ \frac{1}{2}b_{l'0}^{T}b_{l'6}^{E}\left( C_{l-2 l' 6}^{L 4}d4(L)+ C_{l2 l' 6}^{L 8}d8(L)\right)  \right]c(l, l', L). 
\end{eqnarray}
We will make use of the simplified form Eq. (\ref{eq: new_All'}) of the bias matrix and estimate the computation time involved in the process. 
The Clebsch-Gordan coefficients can be computed from the Wigner 3jm symbols using the code by \cite{Schulten:1976} written in Fortran 77 based on a recursive evaluation of the 3j coefficients. The Wigner-d functions can be computed using Fourier transforms on the rotation group SO(3). One approach adopted by \cite{Risbo:1996} is  to expand the Wigner-d functions into a Fourier sum and handle the transforms using a trivariate FFT. An alternative approach proposed by \cite{Kostelec:2003} consists to use a recursive evaluation of the Wigner-d functions combined with a bivariate FFT technique. We will use the later through the free software C routine SOFT which calculates the Wigner-d on the rotation group SO(3) with Fourier transforms. Notice that when looping over $L$ in Eq. (\ref{eq: new_All'}), at each step we need to call eight Clebsch coefficients in addition of the Wigner-d functions computed by the formula Eq. (\ref{eq: d2468L}) and Eq. (\ref{eq: dL'm'sym}). Obviously, this is computationally expensive; consequently, the computation on the fly of the Clebsch and Wigner-d functions is not an envisageable option. Our main motivation is to \textit{precompute} \textit{all} Clebsch and Wigner-d coefficients that will allow as to optimize the computation time. We show in this section that the bias matrix can be numerically computed in a very short time without the need of parallel computation. Due to the huge RAM memory requirement (see, \citealt{Kostelec:2003}) for the Wigner-d computation the maximum multipole is limited to $l_{max}=500$. In addition, we know that the bias matrix is not far from diagonal. As a result, we can restrain the computation of the bias matrix to a diagonal band $\vert l- l'\vert\leq 20$ which sufficiently provides an accurate estimation of the uncertainties inherent to the non-circularity of the beam. 

\begin{figure*}
\begin{center}
\includegraphics [scale=0.8]{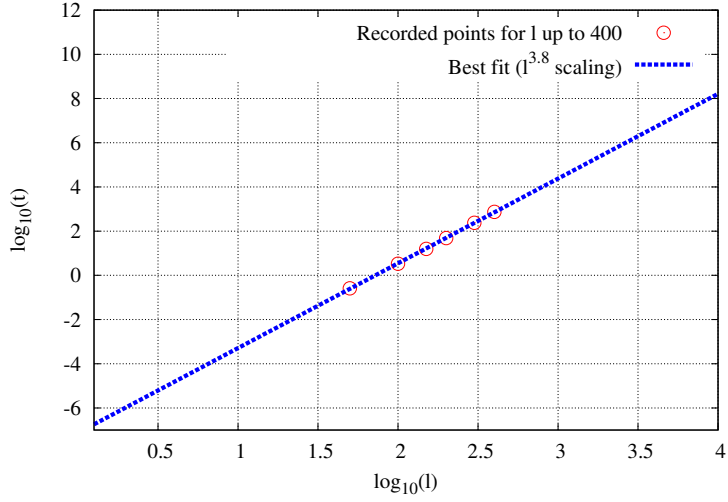}
\caption{ The figure illustrates the plot of the logarithm of the computation time \textit{t} of the sum $\sum_{m=-l}^{l}C_{l-m l' m}^{L 0}$ in minutes as a function of the logarithm of the multipole $l$. The red circled data points are the measured computation time with $l_{max}=400$. The blue dashed line is the best fit to the points. We can see a tight correlation between $\log_{10}(\rm computation\ time)$ and $\log_{10}(\rm l)$. The linear correlation allows us to extrapolate the computation time for large multipoles. The figure indicates a computational cost of $\sim O(l^{3.8})$.}
\label{Fig: sumCl}
\end{center}
\end{figure*} 

All computations have been carried out using a 2.53 GHz Intel Core i5 processor with 4 GB of RAM.  First, we precompute the coefficient $c(l, l', L)$ defined by Eq. (\ref{eq: cll'Lsym}). We have uttered the necessity of the precomputation of the Clebsch coefficients, but if we look at the coefficient $C_{l-m l' m}^{L 0}$ where for each $l$ and $l'$, $L$ varies from $\vert l-l'\vert$ to $l+l'$ and $m$ varies from $1$ to $min(l,\ l')$,  we realize that this cannot be achieved due to the enormous memory storage requirement and this is prohibitive even using high performance computing. Alternatively, we can reduce the memory storage by precomputing the sum $\sum_{m=-l}^{l}C_{l-m l' m}^{L 0}$ and calculating the coefficients $C_{l-m l' m}^{L 0}$ on the fly. We report in Fig. \ref{Fig: sumCl} the computation time of the sum $\sum_{m=-l}^{l}C_{l-m l' m}^{L 0}$ as a function of the multipole. The straight line $\log_{10}(\rm time)=3.8\log_{10}(\rm l)- 7.13$ represents the best fit of the recorded data points. Clearly, the computation time (in minutes) scales as $l^{3.8}$ and by extrapolating to higher l's we find that for the \textit{Planck}-like high resolution experiment the time needed for the precomputation of the summation $\sum_{m=-l}^{l}C_{l-m l' m}^{L 0}$ corresponding to $l_{max}=3000$ is $\sim 10^{6}$ minutes. This can be carried out in a reasonable time with the current high performance computing facilities. Assuming that we use 1000 dual core processors working at the specified frequency of $\sim$ 2.5 GHz the precomputation of the sum up to $l=3000$ will take about 10 hours.

\begin{figure*}
\begin{center}
\includegraphics [scale=0.6]{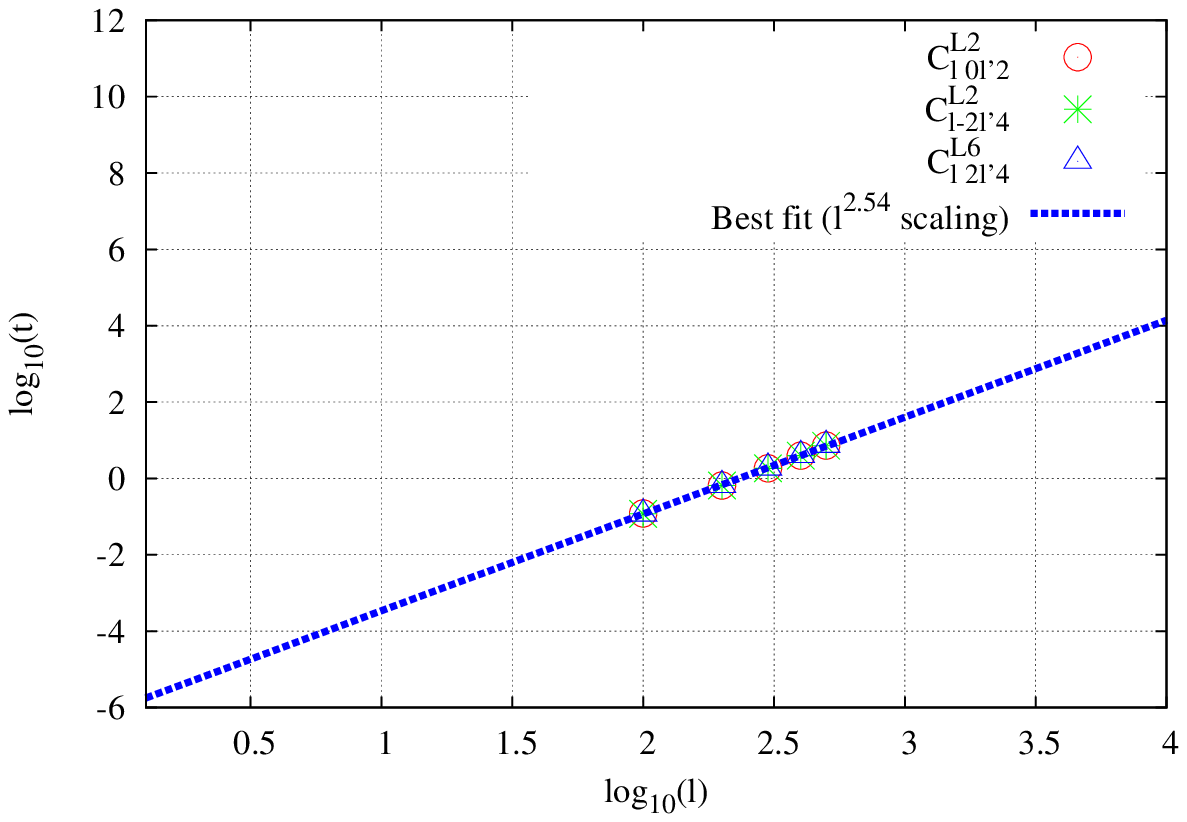}
\includegraphics [scale=0.6]{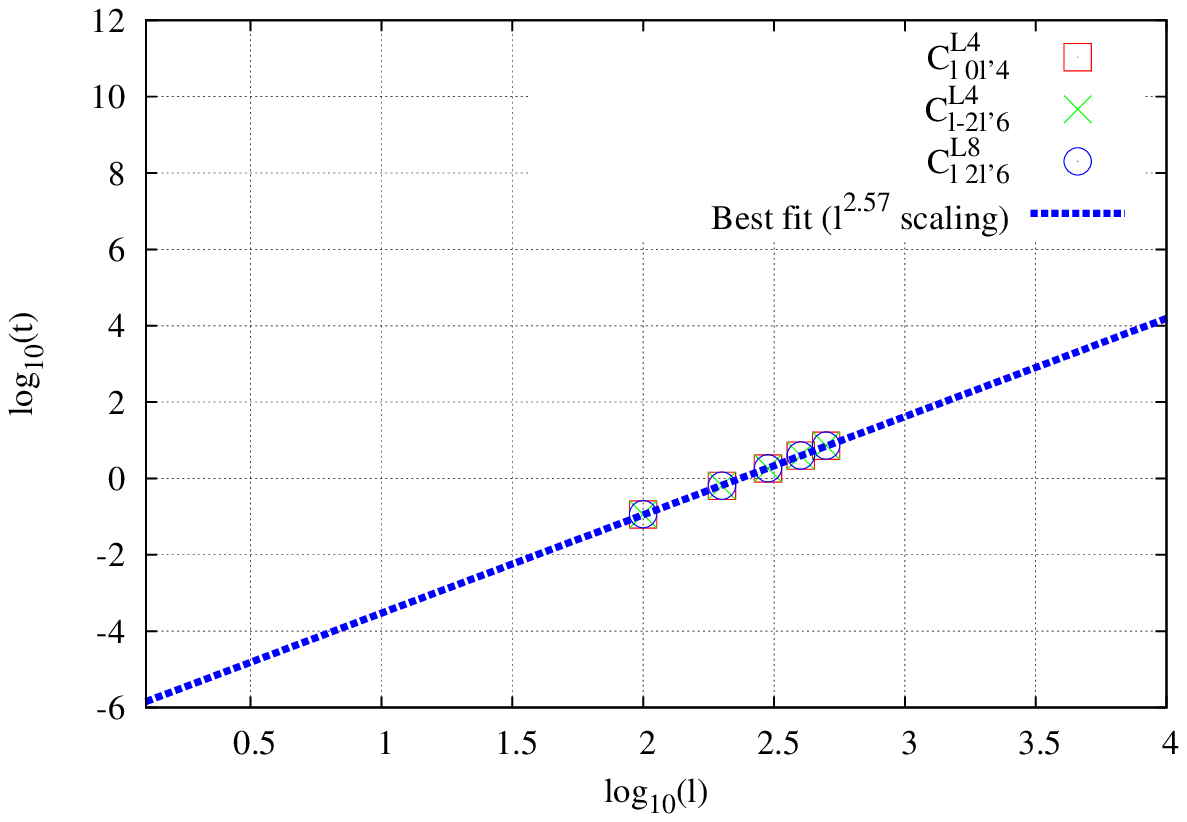}
\caption{The two panels show the computation time (in minutes) of the Clebsch-Gordan coefficients involved in the calculation of the bias matrix. The blue dashed lines in both panels are the best fit of the data points for $l_{max}=500$. Each Clebsch coefficients has approximately the same computation time and scales as $O(l^{2.6})$.}
\label{Fig: Cl}
\end{center}
\end{figure*}

\begin{figure*}
\begin{center}
\includegraphics [scale=0.8]{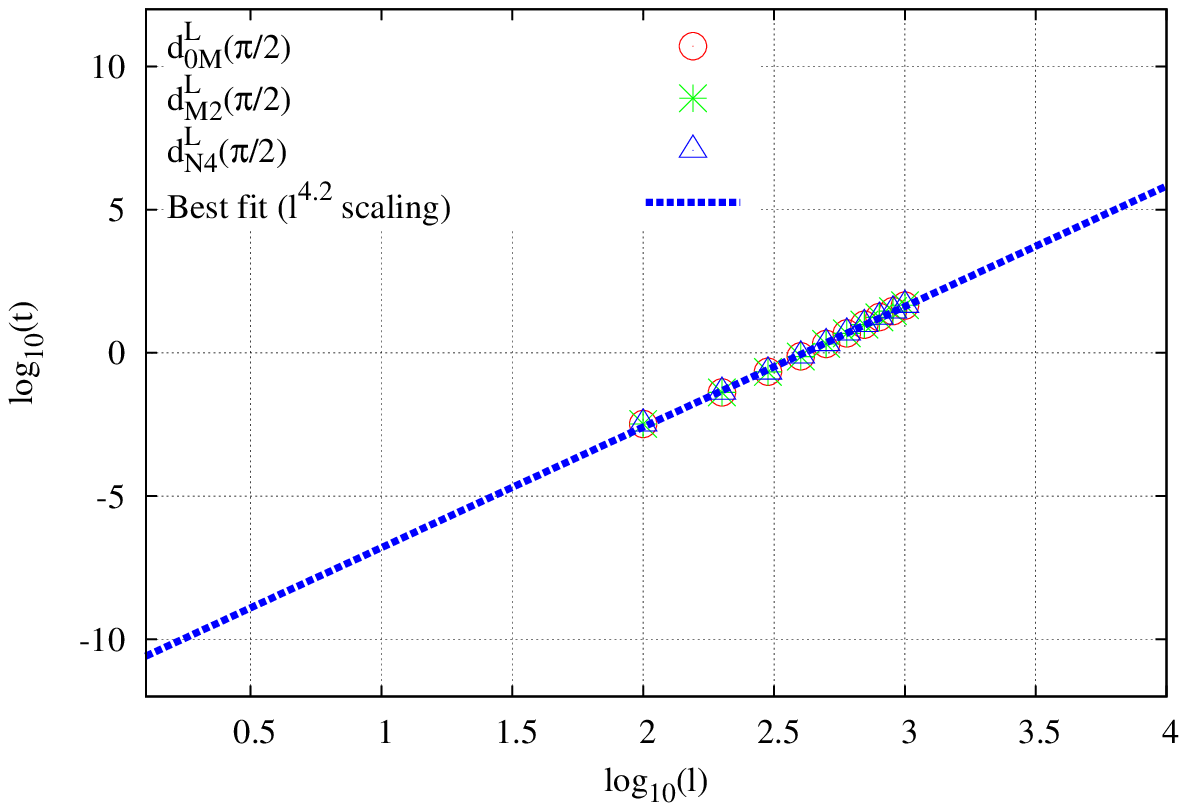}
\caption{The figure illustrates the computation time (in minutes) of the Wigner-d functions for data points recorded up to $l=1000$. The blue dashed line scaling as $l^{4.2}$ represents the best fit to the data points. The slope of the best fit line is relatively much steeper implying that the computation of the Wigner-d function takes much longer time as the multipole $l$ increases.}
\label{Fig: dLM2}
\end{center}
\end{figure*}

We show in Fig. \ref{Fig: Cl} the computation time of each Clebsch coefficient which roughly scales as $l^{2.6}$. The extrapolation of the best fit to $l=3000$ gives an estimate of 12 hours for the precomputation of each Clebsch-Gordan coefficient (1 CPU). Practically, this can be achieved just in a few minutes using computer clusters.
Following the same procedure we precompute all Wigner-d functions of the form $d_{0M}^{L}(\frac{\pi}{2})$ where $1\leq M\leq L$ and $\vert l- l'\vert\leq L\leq l+l'$. The results are shown in Fig. \ref{Fig: dLM2} where the computational cost is $\sim O(l^{4.2})$. A naive estimate of the computation time deduced from the best fit extrapolated to $l=3000$ gives $\sim$3 days for each Wigner-d. We conclude that at sufficiently large multipoles (small scales) the Wigner-d functions dictate the computational complexity of the calculation of the bias matrix. As the Wigner-d  and Clebsch-Gordan coefficients values will never change, we just need to precompute all at once using clusters and store the coefficients in the computer disk.
The next step of the numerical implementation is to precompute the product of the Clebsch coefficients with the terms $dm'(L)$ ($m$=2, 4, 6, 8) introduced in the expression of the bias matrix Eq. (\ref{eq: new_All'}) via Eq. (\ref{eq: d2468L}). We define and precompute the new quantities
\begin{eqnarray}
cd2(l, l',L)&=&C_{l0l'2}^{L2}d2(L)\\
cd4(l, l',L)&=&C_{l0l'4}^{L4}d4(L)\\
cd6(l, l',L)&=&C_{l-2l'4}^{L2}d2(L)+ C_{l2l'4}^{L6}d6(L)\\
cd8(l, l',L)&=&C_{l-2l'6}^{L4}d4(L)+ C_{l2l'6}^{L8}d8(L)
\end{eqnarray}
for each $2\leq l\leq l_{max}=500$, $\vert l- l'\vert \leq 20$, $\vert l- l'\vert \leq L\leq l+l'$. We store the new coefficients in splitted files where each file has 63 MB of size. There are in total 11 output files as we have constrained $l$ and $l'$ to a bandwidth of 20 where $l_{max}=500$. The computation time is shown in Table~\ref{tab: time}. We, again, rearrange the expression of the bias matrix Eq. (\ref{eq: new_All'}) in the following form which is ready for further precomputation 
\begin{eqnarray}
\label{eq: new_All'form1}
 A_{ll'}^{TE}&=&e^{-l^{2}\sigma^{2}}\ \delta_{ll'}\nonumber\\
 &+& \frac{4\pi}{2l+1}\sum_{L=\vert l-l'\vert}^{l+l'} \left[b_{l'2}^{T}b_{l'2}^{E}\ cd2(l, l', L)c(l, l', L)+b_{l'4}^{T}b_{l'2}^{E}\ cd4(l, l', L)c(l, l', L)\right.\nonumber\\
 &+& \left. \frac{1}{2}b_{l'0}^{T}b_{l'4}^{E}\ cd6(l, l', L)c(l, l', L) + \frac{1}{2}b_{l'0}^{T}b_{l'6}^{E}\ cd8(l, l', L)c(l, l', L)\right] . 
\end{eqnarray}
We proceed as previously and precompute the coefficients defined by
\begin{eqnarray}
cdm'c(l, l', L)=cdm'(l, l', L)c(l, l', L)
\end{eqnarray}
where $m'=2,\ 4,\ 6,\ 8$. We report the computation time of the coefficients in Table~\ref{tab: time}. 
\begin{table*}
\caption{Estimate of the total computation time (in minutes) required for the precomputation of all new coefficients introduced in the calculation of the bias matrix where $l_{max}=500$.}
\begin{center}
\begin{tabular}{lcccccccccccc}
\hline \hline
 Coefficients             		&cd2	&cd4	&cd6	&cd8	&cd2c	&cd4c	&cd6c	&cd8c	&$\Sigma_{2}$ 	&$\Sigma_{4}$	&$\Sigma_{6}$	&$\Sigma_{8}$\\ \hline
 Computation time (mn)	&5.05	&6.60	&3.05	&4.58	&5.04	&4.56	&4.55	&4.60	&1.33			&1.33			&1.33			&1.33	     \\
\hline \hline
\label{tab: time}
\end{tabular}
\end{center}
\end{table*}

\begin{figure*}
\begin{center}
\includegraphics [scale=0.8]{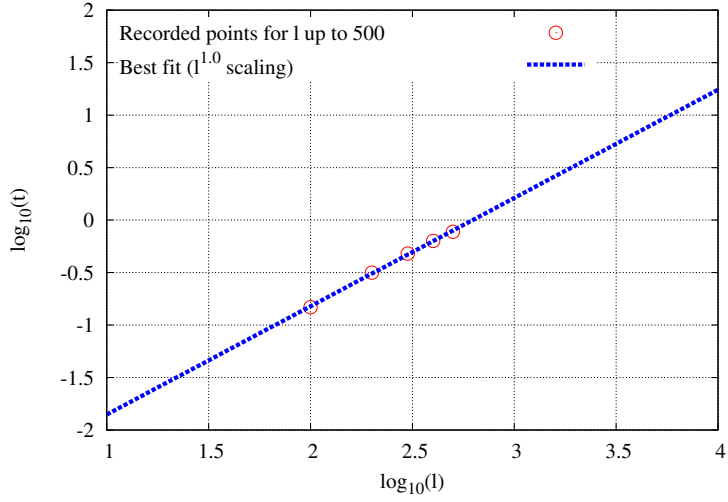}
\caption{ Estimate of the computation time (in seconds) of the bias matrix. The blue dashed line represents the best fit to the red circled data points recorded at $l=100,\ 200,\ 300,\ 400,\ 500$. The total amount of computation time is considerably reduced and scales as $O(l)$. Note that the computation time is in a unit of seconds.}
\label{Fig: Allp_TE_time}
\end{center}
\end{figure*}

We plug in Eq. (\ref{eq: new_All'form1}) the new coefficients and write the formula in the following form
\begin{eqnarray}
\label{eq: new_All'form2}
 A_{ll'}^{TE}&=&e^{-l^{2}\sigma^{2}}\ \delta_{ll'}\nonumber\\
 &+& \frac{4\pi}{2l+1} \left[b_{l'2}^{T}b_{l'2}^{E} \sum_{L=\vert l-l'\vert}^{l+l'}cd2c(l, l', L)+b_{l'4}^{T}b_{l'2}^{E} \sum_{L=\vert l-l'\vert}^{l+l'}cd4c(l, l', L)\right.\nonumber\\
 &+& \left. \frac{1}{2}b_{l'0}^{T}b_{l'4}^{E} \sum_{L=\vert l-l'\vert}^{l+l'}cd6c(l, l', L) + \frac{1}{2}b_{l'0}^{T}b_{l'6}^{E} \sum_{L=\vert l-l'\vert}^{l+l'}cd8c(l, l', L)\right] . 
\end{eqnarray}
The final step of the algorithm implementation consists to precompute the summations defined by
\begin{eqnarray}
\Sigma_{2}(l, l')=\sum_{L=\vert l-l'\vert}^{l+l'}cd2c(l, l', L)\\
\Sigma_{4}(l, l')=\sum_{L=\vert l-l'\vert}^{l+l'}cd4c(l, l', L)\\
\Sigma_{6}(l, l')=\sum_{L=\vert l-l'\vert}^{l+l'}cd6c(l, l', L)\\
\Sigma_{8}(l, l')=\sum_{L=\vert l-l'\vert}^{l+l'}cd8c(l, l', L)
\end{eqnarray}
which only depend on $l$ and $l'$. We resume on Table~\ref{tab: time} the computation time of the different coefficients that we have incorporated in the bias matrix. Then the final form of the bias matrix reads
\begin{eqnarray}
\label{eq: final_All'form}
 A_{ll'}^{TE}&=&e^{-l^{2}\sigma^{2}}\ \delta_{ll'}\nonumber\\
 &+& \frac{4\pi}{2l+1} \left[b_{l'2}^{T}b_{l'2}^{E}\ \Sigma_{2}(l, l')+b_{l'4}^{T}b_{l'2}^{E}\ \Sigma_{4}(l, l')\right.\nonumber\\
 &+& \left. \frac{1}{2}b_{l'0}^{T}b_{l'4}^{E}\ \Sigma_{6}(l, l') + \frac{1}{2}b_{l'0}^{T}b_{l'6}^{E}\ \Sigma_{8}(l, l')\right] . 
\end{eqnarray}
From Eq. (\ref{eq: final_All'form}) we can estimate the power spectrum using the relation defined in Eq. (\ref{eq: C_TE_expectation_value}) and investigate how the beam asymmetry affects the cross-power spectrum TE. As all coefficients have been precomputed, provided the beam width and eccentricity, only the beam harmonic transforms need to be computed using the specific model of beam. The final computational cost of the bias matrix evaluation is illustrated in Fig. \ref{Fig: Allp_TE_time}. We have already noticed that $\log_{10}(\rm computation\ time)$ varies linearly with $\log_{10}(\rm l)$. Therefore, we can fit the data points recorded from the runs with a linear function. The equation of the best fit is given by $\log_{10}(\rm time)=1.03\log_{10}(\rm l)-2.88$. After extrapolation to $l=3000$, we find that the bias matrix can be computed in just 5 seconds. We find that the computational gain of the method scales as $O(l_{max}^{6.0})/O(l_{max}^{4.2})=O(l_{max}^{1.8})$. 

\section{non-circular beam investigations}
\label{Sec: non-circular}
\begin{figure*}
\begin{center}
\includegraphics [scale=0.6]{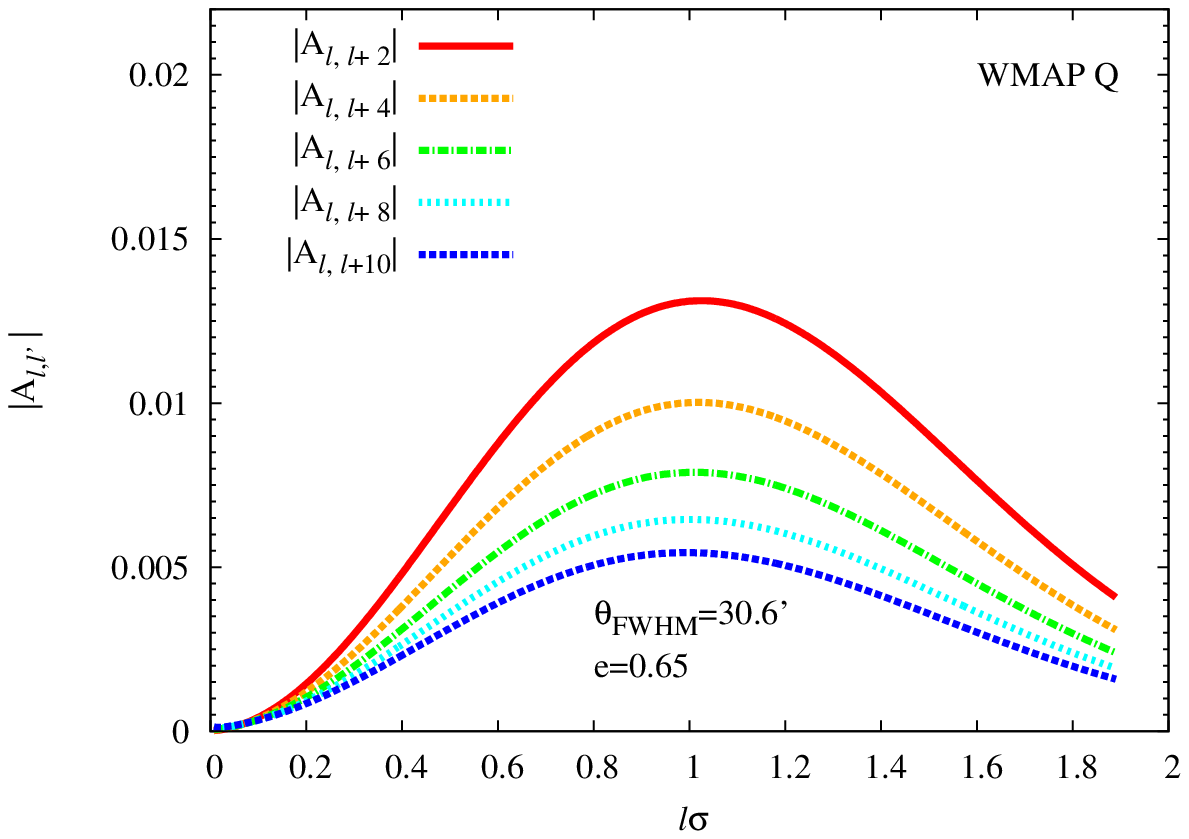}
\includegraphics [scale=0.6]{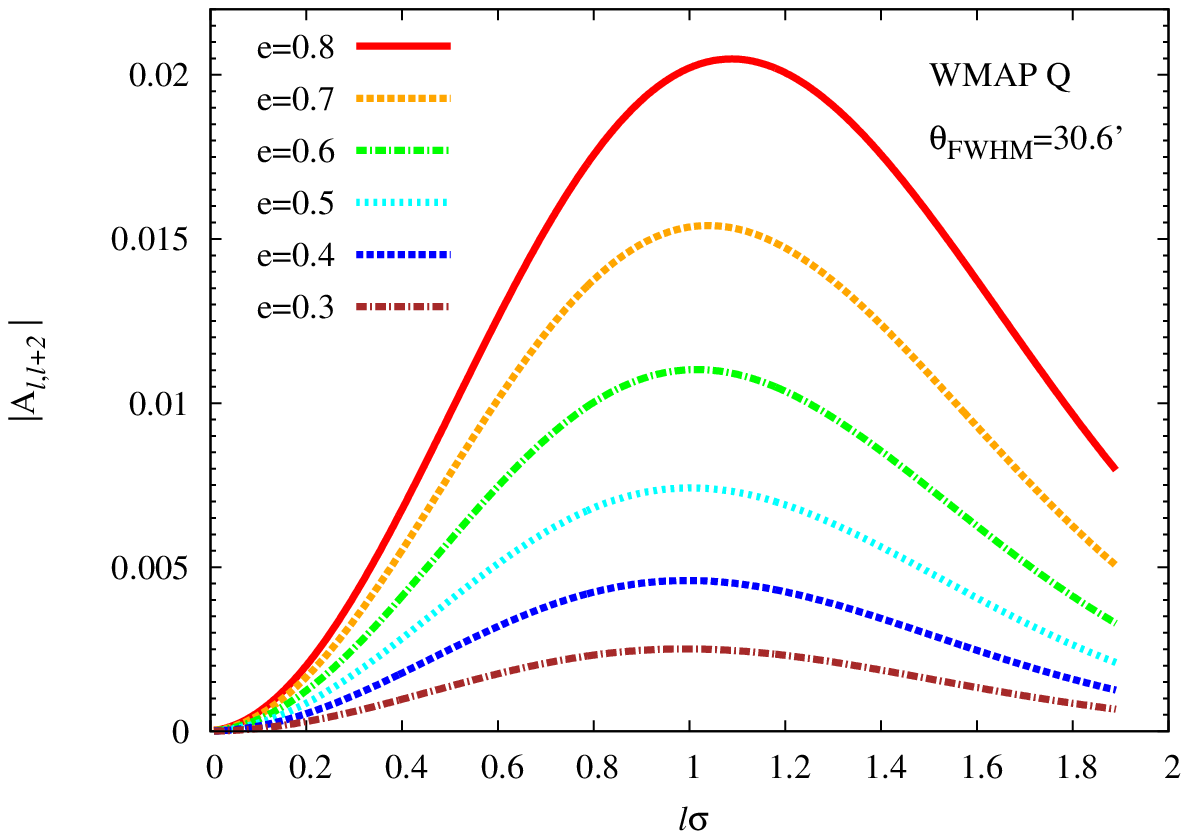}
\caption{Plots of the modulus of the bias matrix as  a function of the multipole $l$, for different values of $l'-l=2,\ 4,\ 6,\ 8,\ 10$. The left panel illustrates the coupling between multipoles arising from the non-circularity of the beam. The mixing of power between multipoles kicks in at $l\sigma \sim$1 but decreases when we move away from the diagonal. We show in the right panel the effect of the beam eccentricities for a given beam size $\theta_{\rm FWHM}=30.6'$ at $l'=l+2$. The bias increases rapidly with the beam ellipticity (eccentricity) whereas the peaks of the bias are shifted to higher $l$s. The elements of the bias matrix are shown for a model of beam corresponding to the \textit{WMAP}-Q1 band with the mean beamwidth $\sigma=3.78\times10^{-3}$.}
\label{Fig: Allplus}
\end{center}
\end{figure*}

In this section we present a detailed analysis of the beam systematics using realistic beam geometry from the \textit{WMAP} and \textit{Planck} experiments. We have reviewed that any deviation of the beam from circularity induces systematic errors in the estimation of the power spectrum which become especially significant at small angular scales. As the multipole $l$ increases we expect more off-diagonal elements in the bias matrix which arise from the non-circularity of the beam. In that case the bias matrix encodes the coupling between multipoles which is illustrated in Fig. \ref{Fig: Allplus} for the \textit{WMAP} experiment in Q1 band at the frequency 40.9 GHz \citep{Page:2003}. As inferred from both panels the bias inherent to the beam asymmetry dominates at $l\sigma\sim 1$. The coupling between multipoles due to the non-circularity of the beam is evident but the effect decreases when we move further away from the diagonal elements of the bias matrix.  The bias systematic is  effectively pronounced for highly elliptical beams. 

\begin{figure*}
\begin{center}
\includegraphics [scale=0.6]{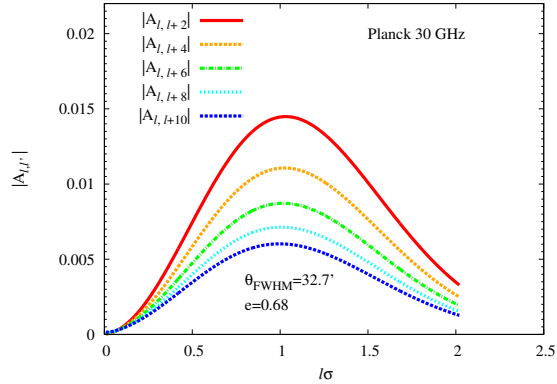}
\caption{The modulus of the bias matrix plotted as  a function of the multipole $l$, for different values of $l'-l=2,\ 4,\ 6,\ 8,\ 10$. The coupling between multipoles arising from the beam asymmetry kicks in at $l\sigma \sim$1 but falls off when we move away from the diagonal.  The elements of the bias matrix are shown for a model of beam corresponding to the \textit{Planck} 30 GHz with the mean beamwidth $\sigma=4.04\times10^{-3}$ and mean ellipticity $\epsilon=1.36$ (\textit{e}=0.68).}
\label{Fig: Allplus_30GHz}
\end{center}
\end{figure*}

In new generation CMB high resolution experiments the beam systematics can affect significantly the estimation of the power spectrum. We consider particularly the \textit{Planck} instrument beam response that can be simulated with the beam model defined in Eq. (\ref{eq: beam_harmonics}). The \textit{Planck} survey is carried out with the Low Frequency Instrument (LFI) \citep{Bersanelli:2010}, and the High Frequency Instrument (HFI) \citep{Lamarre:2010}. The broad frequency range of \textit{Planck} allows to cover the peaks of the CMB power spectrum and characterizes the spectra of foreground emissions \citep{Tauber:2010}. The \textit{Planck} polarized detectors at 30 GHz exhibit the highest beam asymmetry with an ellipticity which spans over the range $1.35-1.40$ \citep{Mitra:2011}. Therefore, we expect an important beam corrections at this channel. At 30 GHz the beam mean ellipticity is $\epsilon=1.36$ (\textit{e}=0.68) with a mean beam width $\theta_{\rm FWHM}=32.7'$ \citep{Tauber:2010}. We report in Fig. \ref{Fig: Allplus_30GHz} the bias matrix obtained from the simulated beams. The plot of the bias matrix against the multipoles exhibits similar behaviour as Fig. \ref{Fig: Allplus} demonstrating the importance of power mixing between multipoles in polarization experiments with asymmetric beam.

\begin{figure*}
\begin{center}
\includegraphics [scale=0.6]{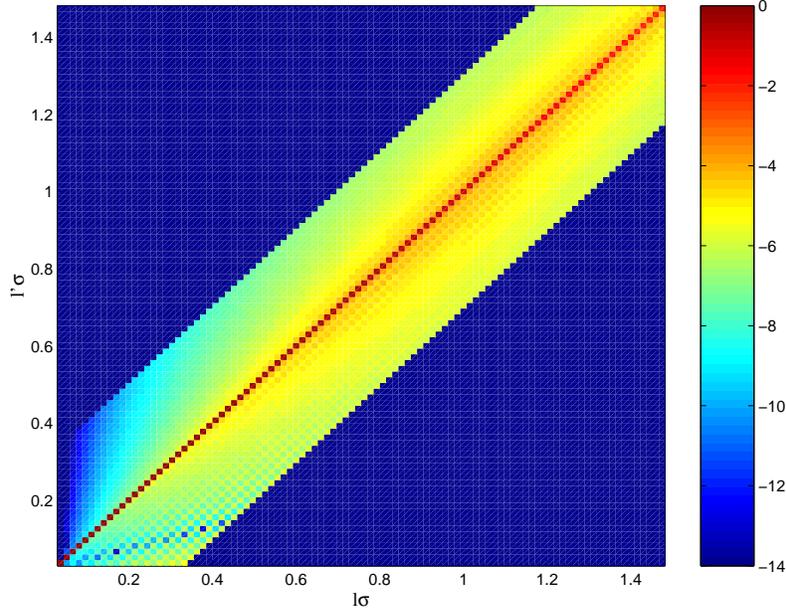}
\caption{Plot of $\log\mid A_{ll'}\mid$ in regions of the multipole space for an hypothetical experiment under a non-rotating beam assumption with a beam resolution $\theta_{\rm FWHM}=2^{\circ}$ (average beam size $\sigma=1.48\times10^{-2}$) and eccentricity \textit{e}=0.6. Significant off-diagonal elements can be seen at $l\sigma \sim$1 and the effects become more important as \textit{l} increases. The plot of the logarithm of the bias matrix modulus is shown within the mutipoles band width $\mid l - l'\mid$=20.}
\label{Fig: Allp_TE120_60}
\end{center}
\end{figure*}

In order to illustrate the effects of the beam non-circularity in multipole space we consider a sufficiently high resolution beam $\theta_{\rm FWHM}=2^{\circ}$ with a mean beam width $\sigma=1.48\times10^{-2}$. We report in Fig. \ref{Fig: Allp_TE120_60} the corresponding plot of the logarithm of the modulus of the bias matrix. We can see that the off-diagonal elements of the bias matrix arising from the non-circularity of the beam start to dominate at the multipole where the bias peaks ($l\sigma=1$).

\begin{figure*}
\begin{center}
\includegraphics [scale=0.6]{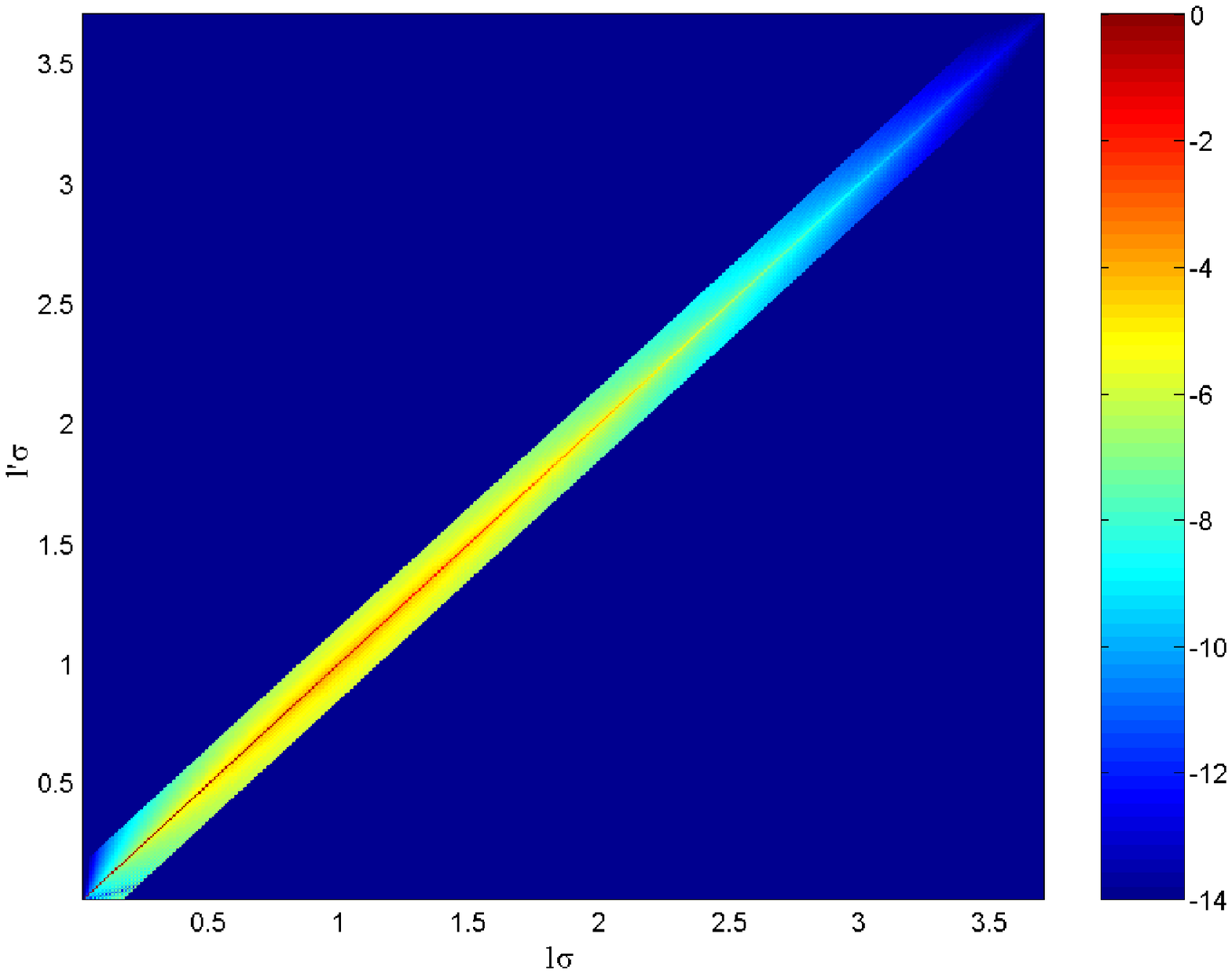}
\includegraphics [scale=0.6]{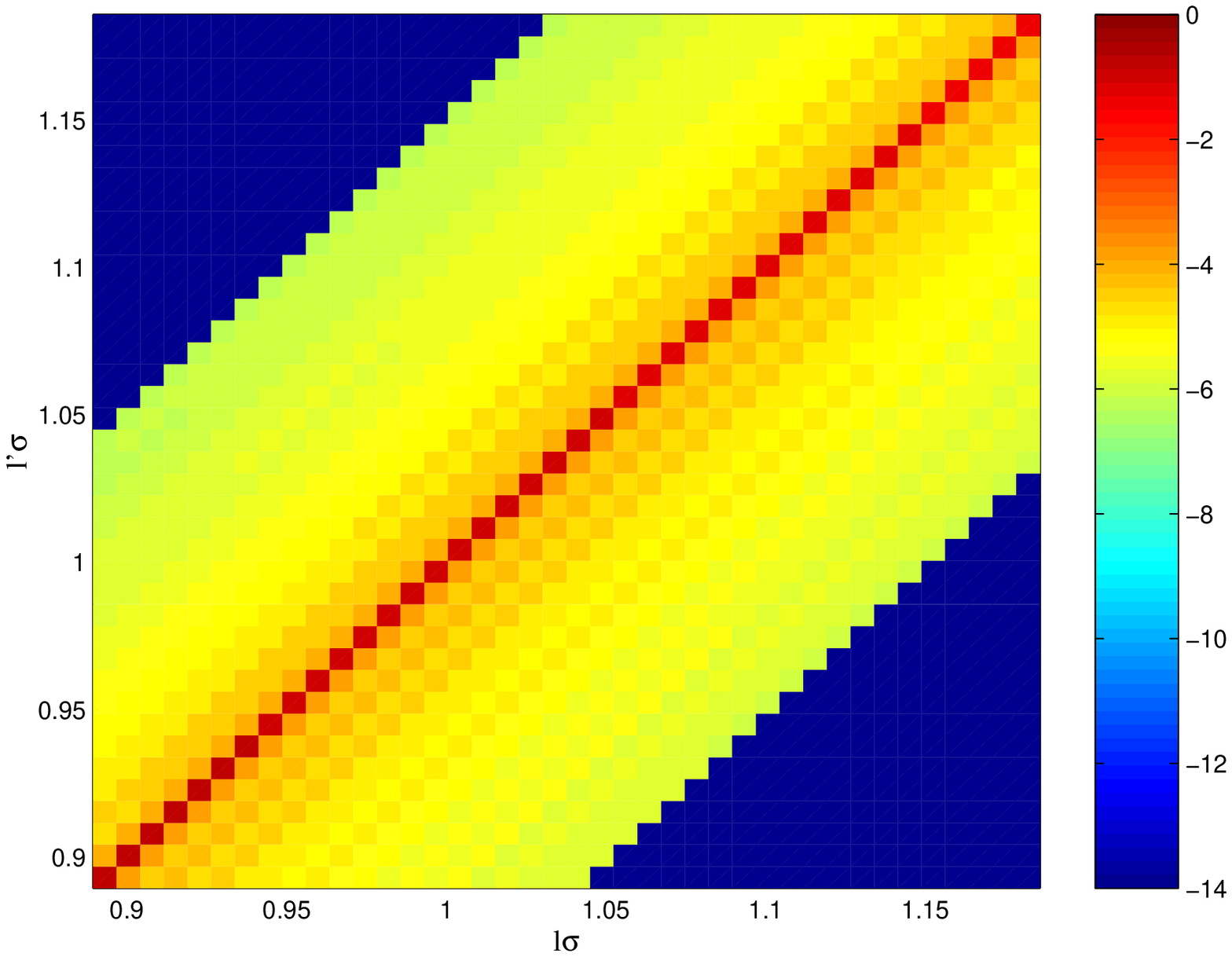}
\caption{Plots of $\log\mid A_{ll'}\mid$ in the multipole space for a hypothetical experiment with non-rotating beam and a beam resolution $\theta_{\rm FWHM}=1^{\circ}$ ($\sigma=7.41\times10^{-3}$) with an eccentricity \textit{e}=0.6. The top panel illustrates the plot between the multipole range [2, 500] and the bottom panel shows the same plot between [120, 140]. Both panels show important off-diagonal elements that kicks in for $l\sigma\geq 1$.}
\label{Fig: Allp_TE60_60}
\end{center}
\end{figure*}

The off-diagonal elements which dominate at $l\sigma \sim 1$ can be clearly distinguished in Fig. \ref{Fig: Allp_TE60_60} where we plot the bias matrix for an ideal experiment with non-rotating beam ( $\theta_{\rm FWHM}=1^{\circ}$ and \textit{e}=0.6). 

\begin{figure*}
\begin{center}
\includegraphics [scale=0.6]{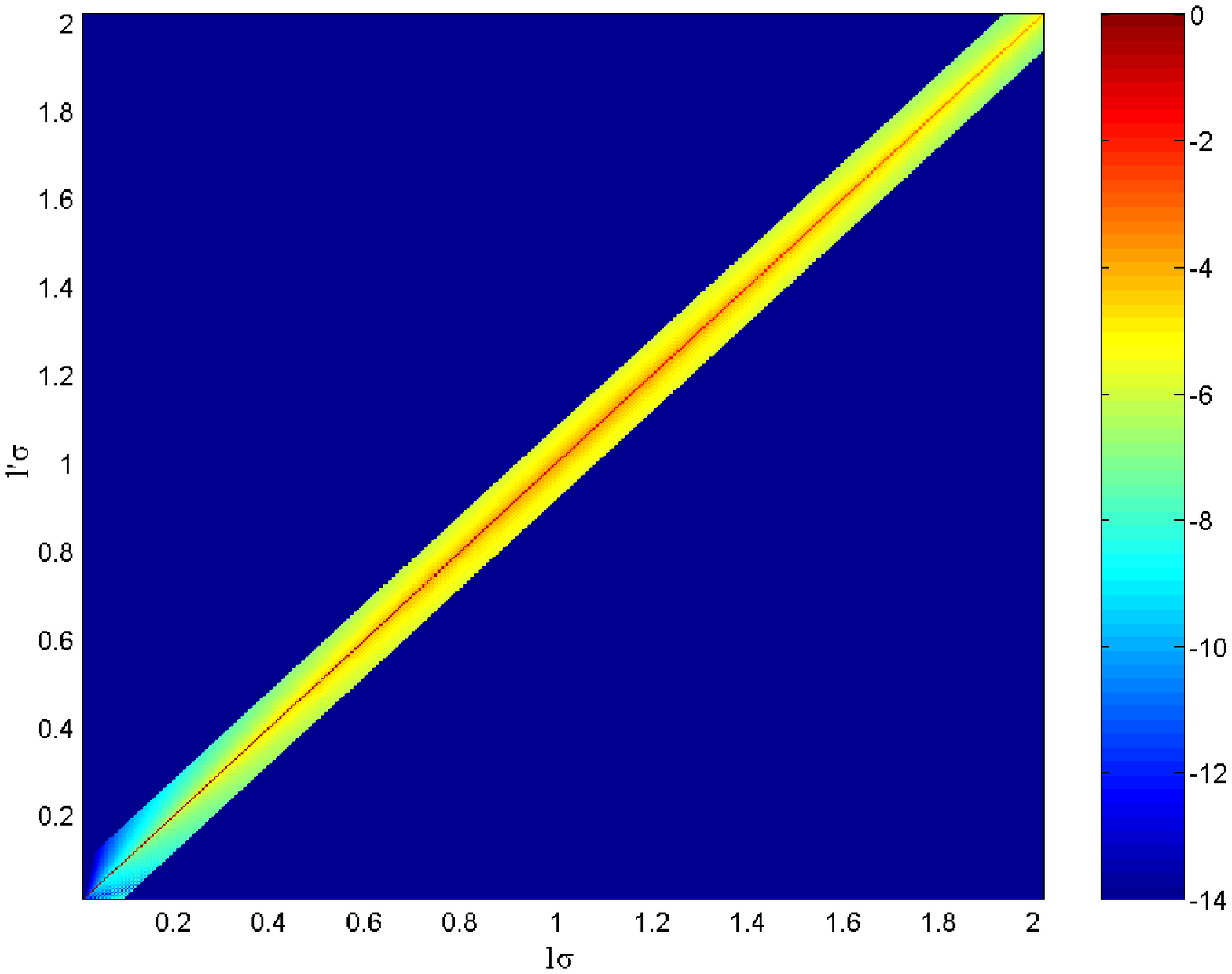}
\includegraphics [scale=0.6]{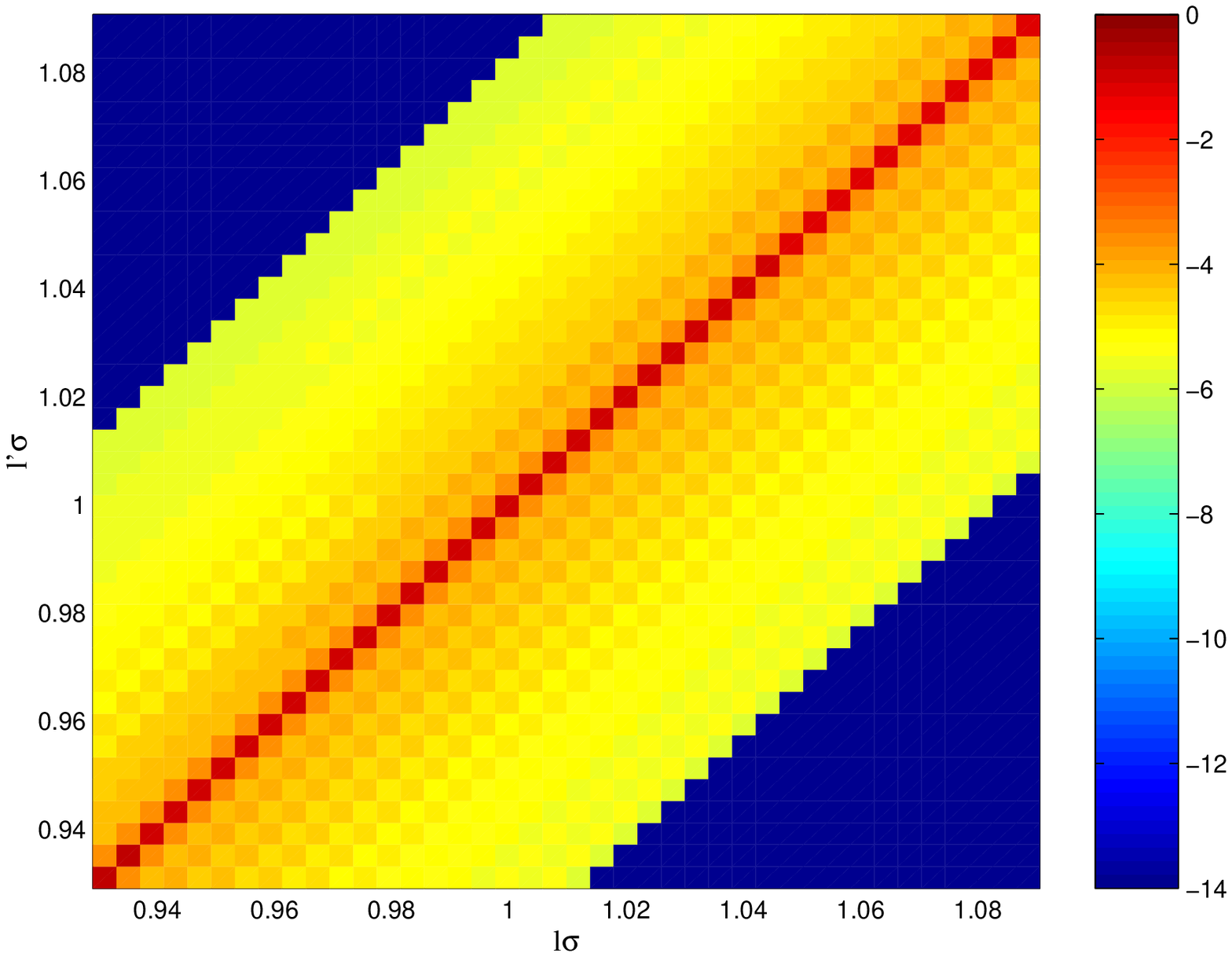}
\caption{Plots of $\log\mid A_{ll'}\mid$ in the multipole space for \textit{Planck} experiment simulated beam parameters at 30 GHz with a beam resolution $\theta_{\rm FWHM}=32.7'$ ($\sigma=4.04\times10^{-3}$) and eccentricity \textit{e}=0.68. The top panel illustrates the plot of the bias matrix between the multipole range [2, 500] and the bottom panel shows the same plot between [230, 270]. Both panels show significant mixing of power (off-diagonal elements) between multipoles for $l\sigma\geq 1$ that arises from the beam ellipticity.}
\label{Fig: Allp_TE32_68}
\end{center}
\end{figure*}

We have seen that the \textit{Planck} 30 GHz beam response pattern is the most asymmetric (\textit{e}=0.68) among the \textit{Planck} beams. As an illustration we plot the corresponding bias in the multipole space which is shown in Fig. \ref{Fig: Allp_TE32_68}. Obviously, the coupling between multipoles (off-diagonal elements of the bias) is important for $l\sigma\gtrsim 1$ implying the necessity of an appropriate corrections of the systematics. 

\begin{figure*}
\begin{center}
\includegraphics [scale=0.6]{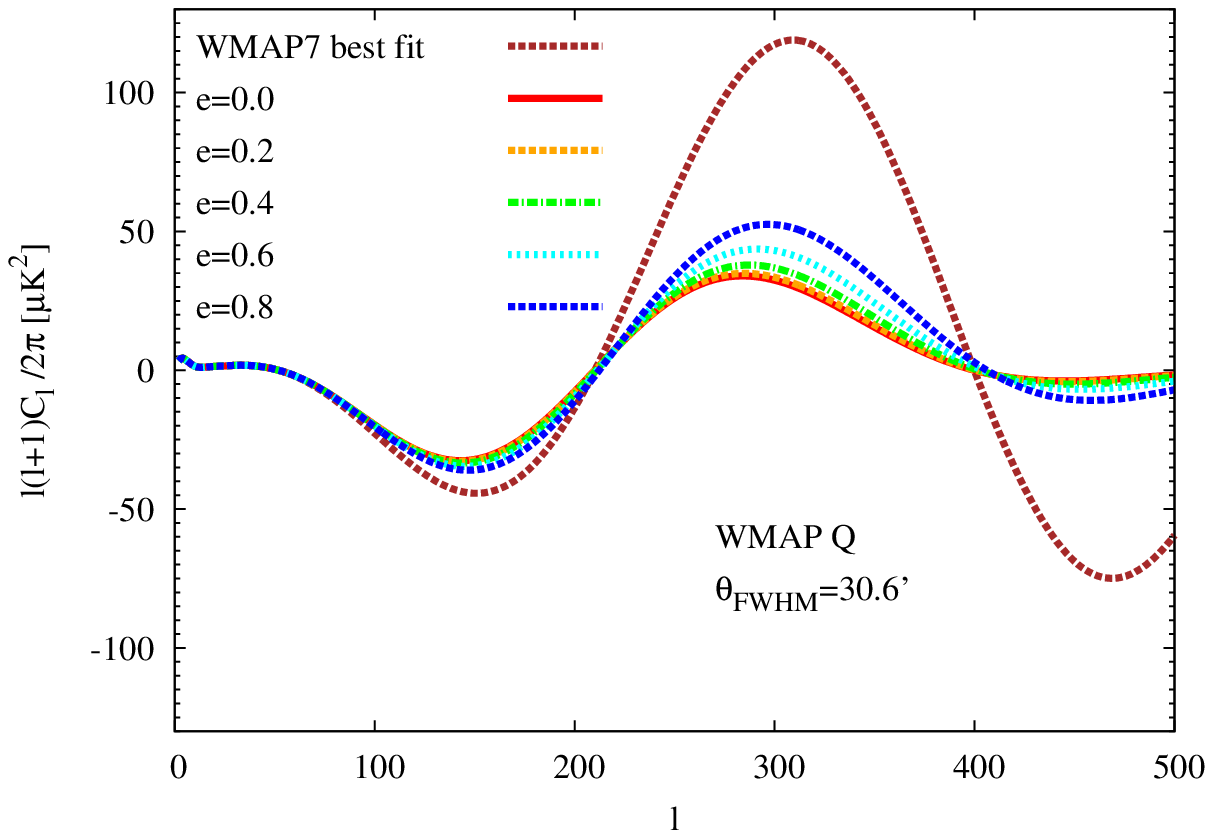}
\includegraphics [scale=0.6]{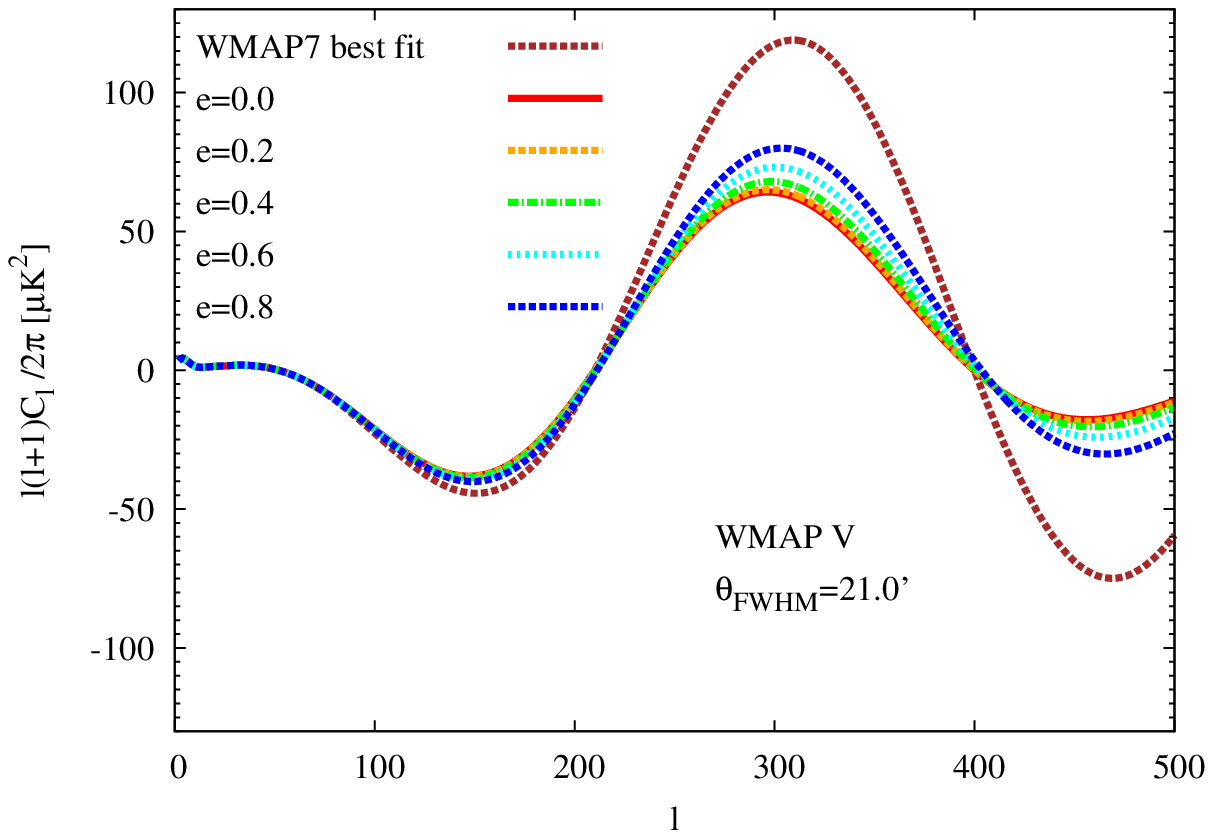}
\caption{Recovered power spectrum using the WMAP7 $\Lambda$CDM  best fit model. The effect of the non-circularity is shown for the \textit{WMAP}-Q1 band experimental beam parameter with size $\theta_{\rm FWHM}=30.6'$ (left panel) and the \textit{WMAP} V band with size $\theta_{\rm FWHM}=21.0'$ (right panel). The red line is the power spectrum computed from a circular Gaussian window. The \textit{WMAP} best fit model is shown in brown dotted line.}
\label{Fig: Cl30}
\end{center}
\end{figure*}

\begin{figure*}
\begin{center}
\includegraphics [scale=0.6]{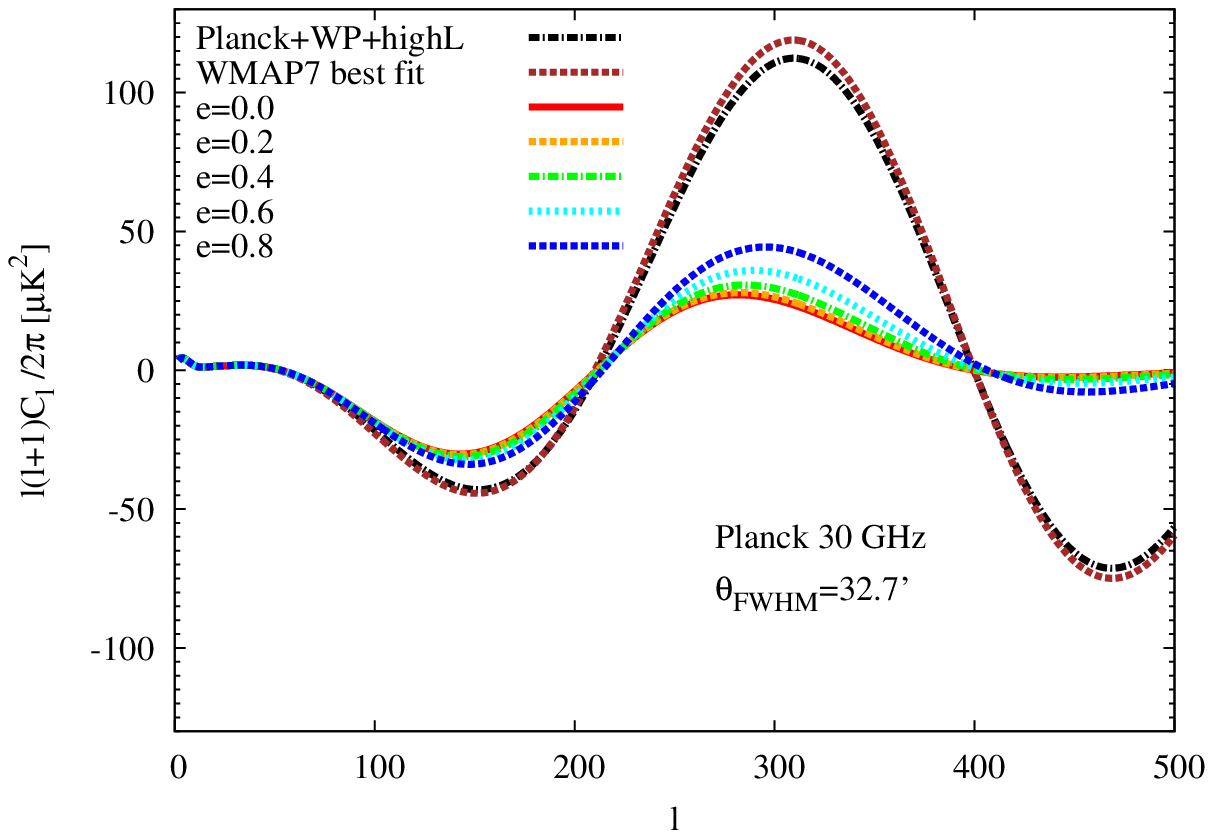}
\includegraphics [scale=0.6]{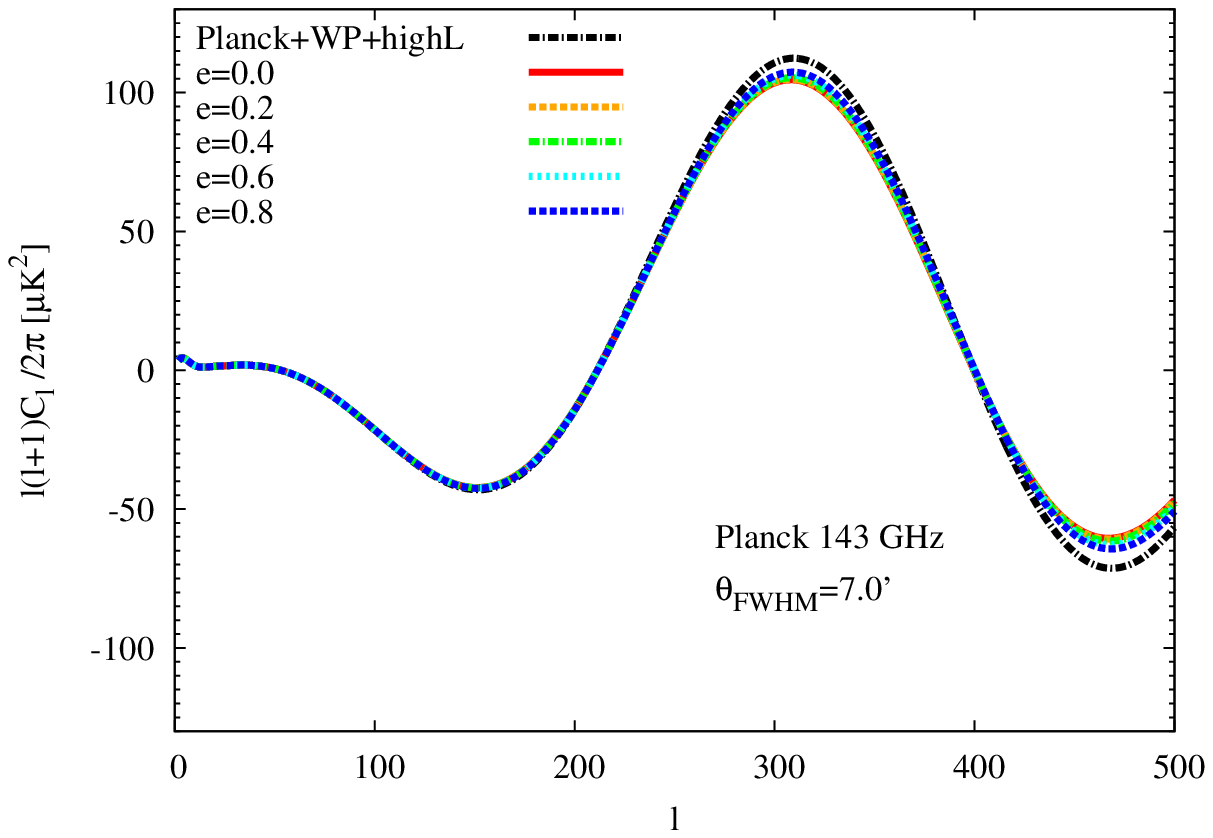}
\caption{Recovered power spectrum using the \textit{Planck} $\Lambda$CDM  best fit model (\textit{Planck}+ \textit{WP}+ highL in black dotted line). For comparison the WMAP7 best fit model is also shown in the left panel (brown dotted line). The effect of the non-circularity is shown for the \textit{Planck} 30 GHz  beam parameter with size $\theta_{\rm FWHM}=32.7'$ and eccentricity \textit{e}=0.68 (highest ellipticity $\epsilon=1.36$) (left panel) and the \textit{Planck} 143 GHz beam with size $\theta_{\rm FWHM}=7.0'$ and eccentricity \textit{e}=0.30 (smallest ellipticity $\epsilon=1.05$) (right panel). }
\label{Fig: Cl32}
\end{center}
\end{figure*}

In the following, we focus our analysis on the effect of the beam asymmetry on the power spectrum estimation and evaluate the uncertainties. For this purpose, we compare the observed power spectrum of an elliptical beam with a given resolution  $\theta_{\rm FWHM}$ and eccentricity \textit{e} to the corresponding power spectrum measured using a circular Gaussian beam with the same size $\theta_{\rm FWHM}$. The observed power spectrum $\langle\tilde C^{TE}_{l}\rangle$ can be obtained by convolving the true power spectrum $C_{l}^{TE}$ with the elliptical window through Eq. (\ref{eq: C_TE_expectation_value}). We compute the true power spectrum using the CAMB \citep{Lewis:2011} software for a set of cosmological parameters derived from the WMAP7 and the \textit{Planck} best fit fiducial model. The recovered power spectrum is illustrated in Fig. \ref{Fig: Cl30}. For a given beam size we find that the peak of the power spectrum is increasing with the beam eccentricity (ellipticity) and is shifted to higher \textit{l}.
Similar shifts are observed for the \textit{Planck}+ \textit{WP}+ highL \citep{Planck:2013} best fit model which are reported in Fig. \ref{Fig: Cl32} for the \textit{Planck} channels with the highest beam asymmetry (\textit{e}= 0.68 at LFI 30 GHz) and the smallest beam asymmetry (\textit{e}=0.30 at HFI 143 GHz).  From both panels of Fig. \ref{Fig: Cl30} and  Fig. \ref{Fig: Cl32} we notice that for a given beam eccentricity the peaks amplitude of the power spectrum increases as the beam size becomes much smaller.

\begin{figure*}
\begin{center}
\includegraphics [scale=0.6]{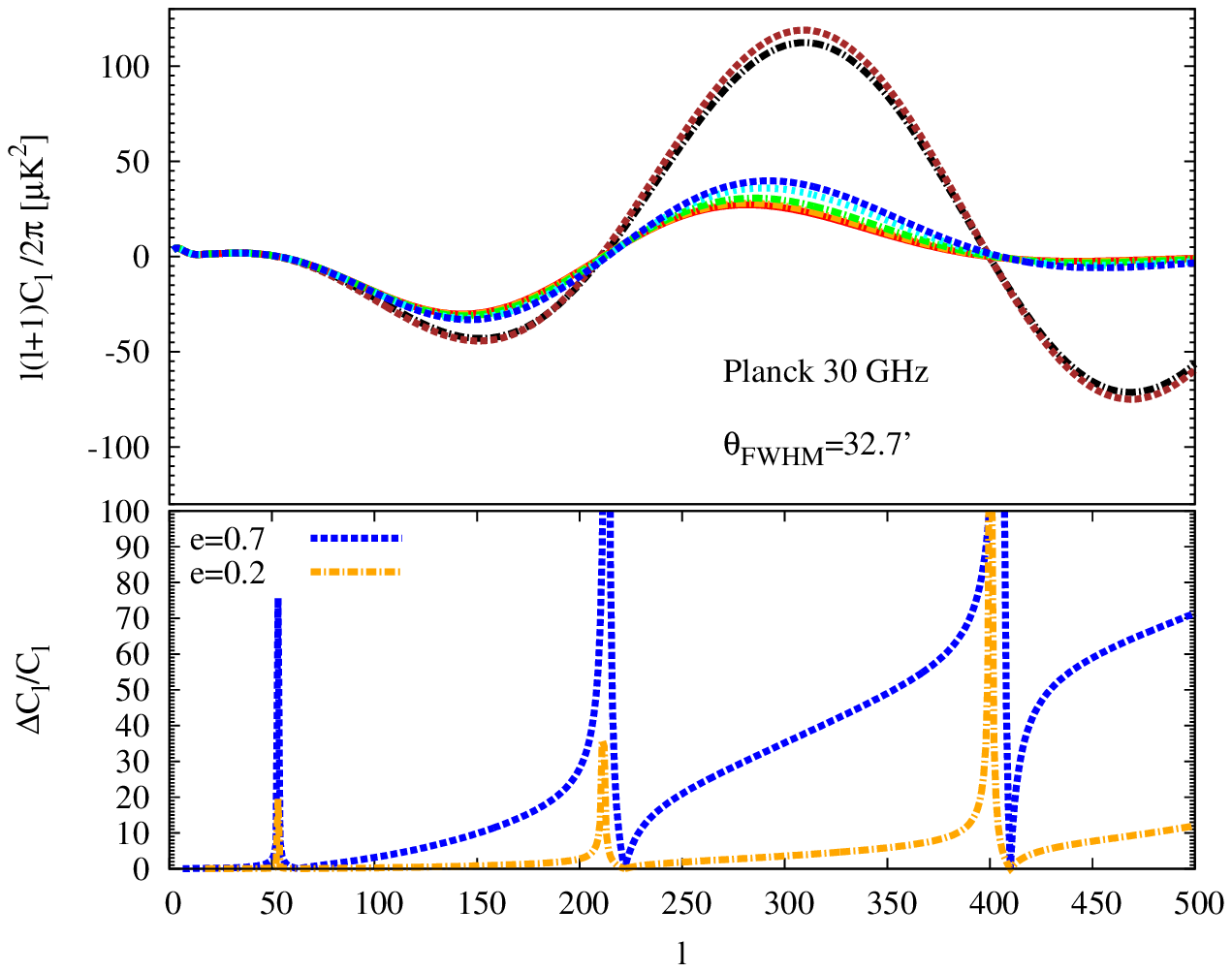}
\includegraphics [scale=0.6]{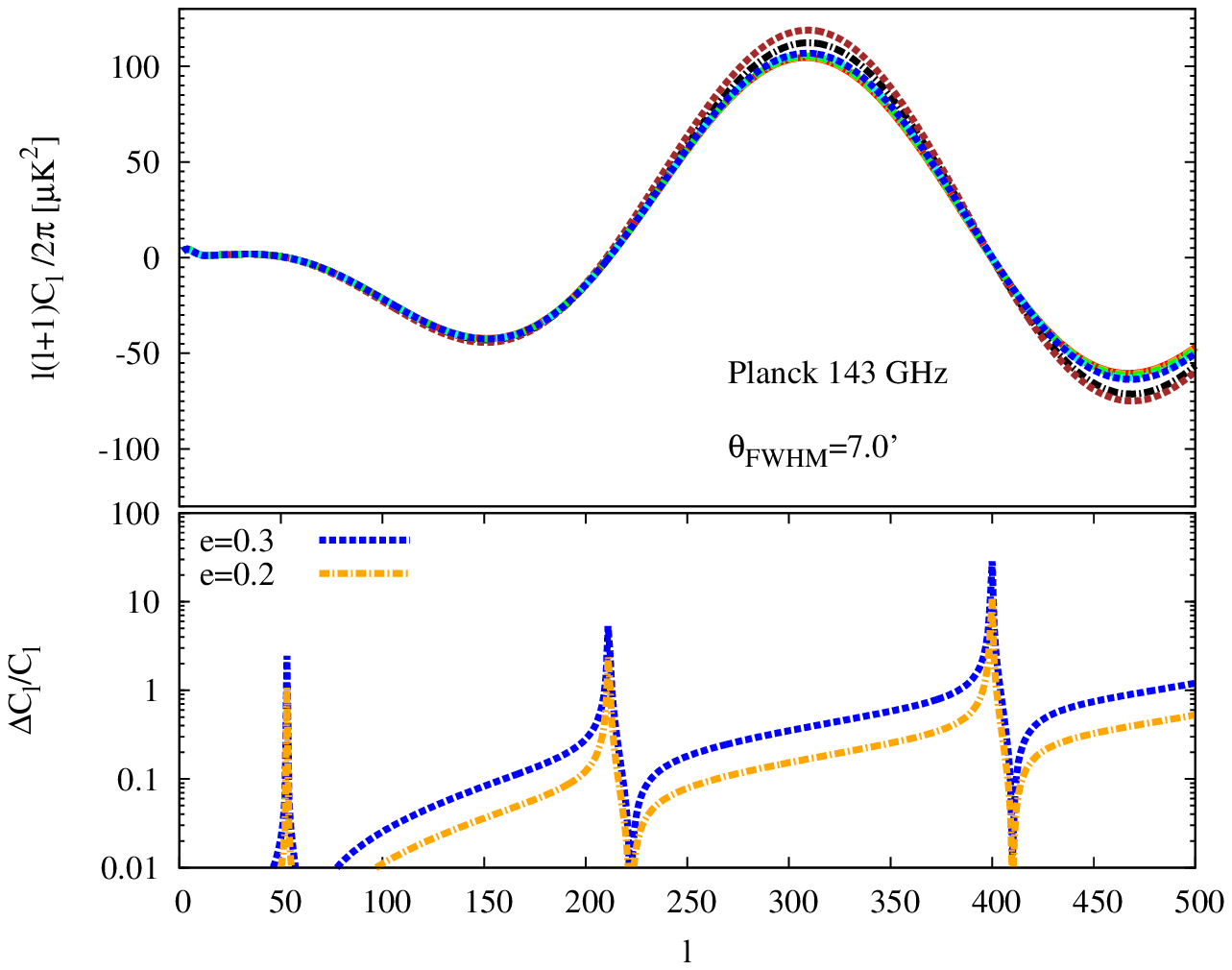}
\caption{Recovered power spectrum using the \textit{Planck} $\Lambda$CDM  best fit model and the corresponding relative uncertainties $\Delta \rm C_{l}/C_{l}$ as a function of the multipole. The systematic errors due to the non-circularity of the beam is shown for the  \textit{Planck} beam $\theta_{\rm FWHM}=32.7'$ and eccentricity $e=0.68\sim 0.7$ at 30 GHz (top panel) and the \textit{Planck} beam $\theta_{\rm FWHM}=7.0'$ and eccentricity $e=0.3$ at 143 GHz (bottom panel). Hypothetical experiments with respectively the same beam size $\theta_{\rm FWHM}=32.7'$ and $\theta_{\rm FWHM}=7.0'$ but with mildly circular beam ($e=0.2$) are shown for comparison. The uncertainties strongly depend on the beam eccentricity and are more significant at small angular scales.}
\label{Fig: Cl32multiplot}
\end{center}
\end{figure*}

The relative uncertainties of the power spectrum estimation are reported in Fig. \ref{Fig: Cl32multiplot}. For clarity we only plotted the systematic errors for the eccentricity $e=0.2$, $e=0.3$ (\textit{Planck} smallest asymmetric beam) and $e=0.7$ (\textit{Planck} highest asymmetric beam). The plots show that the error estimates are significant at large $l$. For the least asymmetric beam (143 GHz) the uncertainties is $\sim$ 1\% at $l_{max}=500$. For the \textit{Planck} 30 GHz the systematic errors can be quite large. However, the uncertainties computed are in reality an upper limit of the systematics since we expect that the effective eccentricity of the beam in the time stream is reduced as during observations the beam revisits each sky pixel with different orientations so that some non-circular modes cancel out.  

\begin{figure*}
\begin{center}
\includegraphics [scale=0.8]{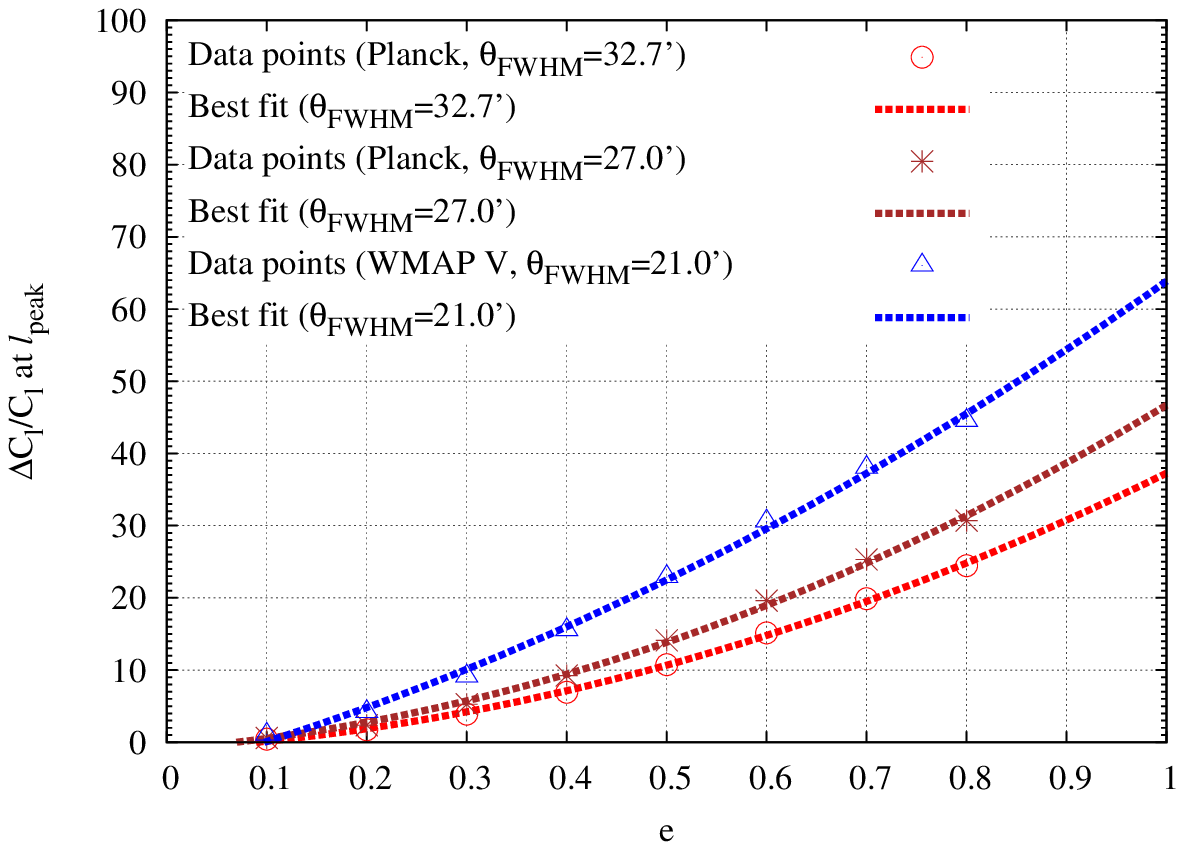}
\caption{Relative uncertainties $\Delta \rm C_{l}/C_{l}$ computed at $l_{\rm peak}=1/\sigma$ for different beam eccentricities \textit{e}. The dashed lines are the best fit to the data points recorded for the \textit{Planck} beam size $\theta_{\rm FWHM}=32.7'$ at 30 GHz, and $\theta_{\rm FWHM}=27.0'$ at 44 GHz. The \textit{WMAP} V channel beam width $\theta_{\rm FWHM}=21.0'$ is shown as well in order to illustrate the effect of the beam size. The data points can be well fitted with a second order polynomial of the eccentricity. As inferred from the plots, the beam systematics increase with the ellipticity (eccentricity) of the asymmetric beam; and given the ellipticity of the beam, the uncertainties of the power spectrum determination become more significant for smaller beam width.}
\label{Fig: Errors_peak32}
\end{center}
\end{figure*}

Now, we estimate the power spectrum uncertainties for different eccentricities computed at the multipole where the bias peaks ($l_{peak}\sigma=1$) for the \textit{Planck} angular beam size $\theta_{\rm FWHM}=32.7'$ (30 GHz), $\theta_{\rm FWHM}=27.0'$ (44 GHz) and \textit{WMAP} V beam size $\theta_{\rm FWHM}=21.0'$. The results are presented in Fig. \ref{Fig: Errors_peak32}. We find an evidence of strong correlation between the relative errors and the eccentricity of the beam computed at $l_{peak}$. The recorded data points can be fitted very well with quadratic polynomials which allow the determination of the uncertainties at $l_{peak}$ up to the smallest angular resolution probed by \textit{Planck} ($\theta_{\rm FWHM}=4.3'$ at HFI 857 GHz). The plots clearly show that for a polarimetry experiment the power spectrum uncertainties at $l_{peak}$ increase with the beam ellipticity and for a given beam eccentricity (ellipticity) the systematic errors become more significant for smaller beam size.

\section{Conclusion}
\label{Sec: conclusion}
In CMB experiments the bias matrix relates the observed power spectrum to their true values. Among the systematic biases affecting the estimation of the power spectrum, the beam non-circularity (asymmetry) is one of the major potential source of systematic errors that cannot be neglected in \textit{Planck}-like high sensitivity and resolution experiment.  More importantly for the polarized signal (at a level about tenth of the temperature fluctuation), the beam systematics must be correctly addressed and accounted for. We have presented a semi-analytical framework to compute the bias matrix of the TE power spectrum including the beam asymmetry in which the non-circular beam shape was modeled using a perturbative expansion of the beam around a circular Gaussian beam (see, \citealt{Fosalba:2002}). \\
We have developed a computationally fast algorithm that can be implemented in simulation pipeline analysis although the formalism described in this paper is only valid for a non-rotating beam. Nevertheless our approach provides some insights about the computational cost involved in more realistic scanning strategies. We have reduced the analytical expression of the bias matrix to its simplest form by precomputing all Clebsch-Gordan coefficients and Wigner-d functions. The computation ressource requirement is very modest and can be carried out with  a single computer processor up to the multipole $l_{max}=500$ with a computational scaling of $O(l_{max})$ (1 CPU at 2.53 GHz and 4 GB of RAM). Our approach can be extended to higher multipoles up to $l_{max}=3000$ at the expense of disc/memory storage and code input/output (I/O) overhead. This can be done in the future provided that all Wigner-d functions can be precomputed up to $l_{max}=3000$. The bias matrix computation was restricted in the band $\mid l- l' \mid \leq 20$ in order to reach the highest multipole $l_{max}=500$ probed. This band width choice can be well justified as in most CMB experiments the beam profile is mildly elliptical ($\epsilon \leq 1.2$,  \citealt{Fosalba:2002})  implying that the bias matrix is not far from diagonal. For highly asymmetric beams, a much broader band (e.g $\mid l-l'\mid\leq 50$) is desirable in order to evaluate correctly the error estimates of the power spectrum with high precision. \\
A second order expansion in the ellipticity parameter introduced in the beam harmonic transform is accounted for highly elliptical beam (e.g., \textit{Planck} 30 GHz where $\epsilon=1.36$), and two non-circular corrections with modes $m=2,\ 4$ (for the temperature) and $m=4,\ 6$ (for the E-mode polarization)  are included to describe accurately the beam geometry (see, \citealt{Fosalba:2002}). From the extrapolation of the recorded runs, we find that the bias matrix can be computed up to $l_{max}=3000$ in few seconds and the corresponding computational gain is $\sim O(l^{1.8})$. Theoretically, this is a naive estimate of the computation time at large multipoles as the program I/O overhead marginally increases the computation time.\\
Our findings suggest that the systematic biases peak at a multipole comparable to the inverse of the beam width. The amplitude of the peaks of the bias increases with the ellipticity and is more significant at higher multipoles. Similar behaviour has been already observed for the temperature correlation (see, \citealt{Mitra:2004}). A graphic representation of the bias matrix in multipole space shows the importance of multipoles coupling at $l\sigma \gtrsim 1$ which arises from the beam asymmetry but the mixing of power between multipoles falls off when we move away from the diagonal.\\
We find that the effect of the non-circularity of the beam in the TE power spectrum uncertainties at $l_{peak}\sim 1/\sigma$ is important attaining $\sim 20\%$ ($\epsilon_{\rm mean}=1.36$ at 30 GHz)  in the \textit{Planck} highest asymmetric beams. In \textit{WMAP} Q band (effective ellipticity $\epsilon=1.15$ with beam width $\theta_{\rm FWHM}=30.6'$) the power spectrum uncertainty estimate is $\sim 12 \%$  and in \textit{WMAP} V (effective ellipticity $\epsilon=1.09$ and beam width $\theta_{\rm FWHM}=21'$) the corresponding relative error is $\sim 16 \%$. For \textit{Planck} and \textit{WMAP} beams with much smaller size the bias peaks outside the range [2, 500] but we expect much significant error bars for such high resolution beams. We note that the errors previously estimated in the simulated \textit{WMAP} and \textit{Planck} beams represent upper limits. This is explained by the fact that the ellipticity (eccentricity) of the beam in our time domain representation of the elliptical window is slightly larger than the effective beam ellipticity (eccentricity) in the pixel domain since in the later, during the satellite observations the beams visit each sky pixel multiple times, but with different orientations of the beams resulting to some extent in the suppression of non-circular modes. Consequently, the effective beam ellipticity (eccentricity) in the sky map becomes much smaller than our nominal ellipticity (eccentricity) in the time stream.\\
General scanning strategies must be incorporated in our analysis in the future in order to improve the accuracy of the error estimates of the TE power spectrum. The systematics of instrument noise and sky non-uniformity/cut-sky using asymmetric beams are expected to become important as well in CMB polarization experiments. In the near future, we will extend this work in the case of non-circular beam with incomplete sky coverage resulting from the Galactic foreground emissions and point sources masking.\\
Finally, our fast pipeline implementation provides a very convenient tool in the understandings of the beam systematics corrections of the TE power spectrum at large angular scales where many CMB anomalies have been previously observed by both the \textit{WMAP} and \textit{Planck} experiments.
\section*{Acknowledgements}
F. R. would like to acknowledge the South African Square Kilometre Array Project for financial support. Part of the work was carried out at IUCAA. We thank Moumita Aich for providing the code for computing the Clebsch-Gordan coefficients. Some of the results in this paper have been derived using the CAMB \citep{Lewis:2011} package.

\bibliographystyle{mn2e}
\bibliography{myrefs}

 \appendix
 \section{Consistency checks}
 \label{appendix: checks}
 We will show in this appendix that for the particular case of a circular beam, the general expression of the bias matrix Eq. (\ref{eq: Allprime_fmprimeN}) reduces to the usual form of the window function of a symmetric and co-polar beams in polarization experiments. Only the modes $m''=0$ and $M'=\pm 2$ contribute when the beam is circulary symmetric. Replacing this into Eq. (\ref{eq: Allprime_fmprimeN}) we get
 \begin{eqnarray}
 \label{eq: Bias_finalform}
 A_{ll'}^{TE}&=&\pi b_{l'0}^{T}b_{l'2}^{E}\sum_{m=-l}^{l} \sum_{L=\vert l-l'\vert}^{l+l'}C_{l-m l' m}^{L 0}C_{l 0 l' 0}^{L 0}\sum_{M=-L}^{L}d_{0M}^{L}(\frac{\pi}{2})d_{M0}^{L}(\frac{\pi}{2})f_{0M} \sum_{L'=\vert l-l'\vert}^{l+l'}C_{l-m l' m}^{L' 0}\nonumber \\
 &\times &\left. \left[ C_{l2 l' -2}^{L'0}\sum_{N=-L'}^{L'} d_{0N}^{L'}(\frac{\pi}{2})d_{N0}^{L'}(\frac{\pi}{2}) f_{0N}+ C_{l-2 l' -2}^{L'-4}\sum_{N=-L'}^{L'} d_{0N}^{L'}(\frac{\pi}{2})d_{N-4}^{L'}(\frac{\pi}{2}) f_{-4N}\right. \right.\nonumber\\
 &+&\left. C_{l2 l' 2}^{L'4}\sum_{N=-L'}^{L'} d_{0N}^{L'}(\frac{\pi}{2})d_{N4}^{L'}(\frac{\pi}{2}) f_{4N}+ C_{l-2 l' 2}^{L'0}\sum_{N=-L'}^{L'} d_{0N}^{L'}(\frac{\pi}{2})d_{N0}^{L'}(\frac{\pi}{2}) f_{0N}\right]. 
 \end{eqnarray}
 The integral $\sum_{M=-L}^{L}d_{0M}^{L}(\frac{\pi}{2})d_{M0}^{L}(\frac{\pi}{2})f_{0M}$ can be evaluated from the following relation involving spherical harmonics (Eq. (1) of Section 5.9.1 of \citealt{Varshalovich:1988})
\begin{eqnarray}
 \int_{4\pi} Y_{Lm}^*(\hat{q})d\Omega_{\hat{q}}=\sqrt{4\pi} \delta_{L0}\delta_{m0}\nonumber.
\end{eqnarray}
Then we introduce the relation between rotation matrices and the spherical harmonics using  Eq. (1), Section 4.17 of \cite{Varshalovich:1988}
\begin{eqnarray}
Y_{Lm}^*(\hat{q})=\sqrt{\frac{2L+1}{4\pi}}D_{m0}^L(\hat{q},0),
\end{eqnarray}
and making use of the following formula (Eq. (E14) of \citealt{Mitra:2009}) that relates the rotation matrices to the Wigner-d function
\begin{eqnarray}
D^L_{mm^\prime}\left(\phi,\theta,\rho\right) \ = \ i^{m+m^\prime}
\, e^{-im\phi} \sum_{M=-L}^{L} \left[ (-1)^M \,
d^L_{mM}\left(\frac{\pi}{2}\right) \, e^{iM\theta} \,
d^L_{Mm^\prime}\left(\frac{\pi}{2}\right) \right]
e^{-im^\prime\rho},
\end{eqnarray}
we derive 
\begin{eqnarray}
\label{eq: identify }
\sqrt{\frac{2L+1}{4\pi}} i^m \sum_{M=-L}^L (-1)^M
d^L_{mM}\left(\frac{\pi}{2}\right) d^L_{M0}\left(\frac{\pi}{2}\right) \int e^{-im\phi}d\phi \int e^{iM\theta} \sin\theta
d\theta = \sqrt{4\pi} \delta_{L0}\delta_{m0}
\end{eqnarray}
where $\int e^{-im\phi}d\phi=2\pi\delta_{m0}$. We identify both sides of Eq. (\ref{eq: identify }) and by taking into account the definition 
\begin{eqnarray}
f_{0M}=(-1)^{M}\int_{0}^{\pi}d\theta\sin\theta e^{iM\theta}
\end{eqnarray}
derived from Eq. (\ref{eq: Gamma_m'N}) in the case of non-rotating beam, it follows that
\begin{eqnarray}
\sum_{M=-L}^{L}d_{0M}^{L}(\frac{\pi}{2})d_{M0}^{L}(\frac{\pi}{2})f_{0M}=\dfrac{2}{2L+1}\delta_{L0}.
\end{eqnarray}
This implies that the only non-vanishing terms in Eq. (\ref{eq: Bias_finalform}) are obtained for $L=L'=0$ which yields
\begin{eqnarray}
 A_{ll'}^{TE}=4\pi b_{l'0}^{T}b_{l'2}^{E}\sum_{m=-l}^{l} C_{l-m l' m}^{0 0}C_{l 0 l' 0}^{0 0}C_{l-m l' m}^{0 0}(C_{l2 l' -2}^{00}+C_{l-2 l' 2}^{00})
\end{eqnarray}
since the other terms of the summations containing the Wigner-d function vanish unless $L'\geq 4$. We can further simplify the above equation by using the properties of the Clebsch-Gordan coefficients. From Eq. (1) of Section 8.5.1 and Eq. (11) of Section 8.4.2 of \cite{Varshalovich:1988}
 \begin{eqnarray}
C^{00}_{a\alpha b\beta}&=&(-1)^{a-\alpha}\frac{\delta_{ab}\delta_{\alpha,-\beta}}{\sqrt{2a+1}},\\
C^{c\gamma}_{a\alpha b\beta}&=&(-1)^{a+b-c}C^{c\gamma}_{a -\alpha b -\beta}
\end{eqnarray}
we can derive using a simple algebra the following expression of the bias matrix
\begin{eqnarray}
 A_{ll'}^{TE}&=&\frac{8\pi}{2l+1}b_{l'0}^{T}b_{l'2}^{E}\ \delta_{ll'}\sum_{m=-l}^{l}(C_{l-m l' m}^{0 0})^{2}\nonumber\\
 &=&\frac{8\pi}{2l+1}b_{l'0}^{T}b_{l'2}^{E}\ \delta_{ll'},
\end{eqnarray}
and finally from the analytical definition of the beam harmonic transforms in Eq. (\ref{eq: beam_harmonics}) where we plug in $\chi=0$, we recover
\begin{eqnarray}
A_{ll'}^{TE}=e^{-l^{2}\sigma^{2}}\ \delta_{ll'}
\end{eqnarray}
which is the well-known result of the bias matrix for a symmetric beam (see, \citealt{Ng:1999}; \citealt{Challinor:2000}).

\section{Evaluation of the integrals $\rm I_{1}$, $\rm I_{2}$ and $\rm I_{3}$}
\label{appendix: Integrals}
In this appendix we will give the explicit forms of the integrals by using the properties of spin-$s$ spherical harmonics and Wigner-D functions and their relations to the Clebsch-Gordan coefficients. We use the Eq. (11), Section 5.1.5 and Eq. (1), Section 4.17 of \cite{Varshalovich:1988}
\begin{eqnarray}
Y^{\ast}_{lm}(\hat q)&=&(-1)^{m} Y_{l -m}(\hat q),\\
Y^{\ast}_{lm}(\hat q)&=&\sqrt{\frac{2l+1}{4\pi}}D_{m0}^{l}(\hat q,\rho(\hat q))
\end{eqnarray}
and write the spherical harmonic function in the form
\begin{eqnarray}
Y_{lm}(\hat q)=(-1)^{m}\sqrt{\frac{2l+1}{4\pi}}D_{-m0}^{l}(\hat q,\rho(\hat q)),
\end{eqnarray}
and we get
\begin{eqnarray}
\rm I_{1}&=&(-1)^{m}\sqrt{\frac{2l+1}{4\pi}}\int d\Omega_{\hat q}D_{-m0}^{l}(\hat q,\rho(\hat q))D_{m'm''}^{l'}(\hat q,\rho(\hat q)).
\end{eqnarray}
The product of the two Wigner-D functions is expanded in terms of the Clebsch-Gordan coefficients and can be expressed using Eq. (1), Section 4.6 of \cite{Varshalovich:1988} as follows
\begin{eqnarray}
D_{-m0}^{l}(\hat q,\rho(\hat q))D_{m'm''}^{l'}(\hat q,\rho(\hat q))=\sum_{L=\vert l-l' \vert}^{l+l'}C_{l-m l' m'}^{L(-m+m')}D_{(-m+m')m''}^{L}(\hat q,\rho(\hat q))C_{l0 l' m''}^{Lm''},
\end{eqnarray}
and then the integral $\rm I_{1}$ can be written as
\begin{eqnarray}
\rm I_{1}&=&(-1)^{m}\sqrt{\frac{2l+1}{4\pi}}\sum_{L=\vert l-l' \vert}^{l+l'}C_{l-m l' m'}^{L(-m+m')}C_{l0 l' m''}^{Lm''}\int d\Omega_{\hat q}D_{(-m+m')m''}^{L}(\hat q,\rho(\hat q)).
\end{eqnarray}
We calculate the integrals $\rm I_{2}$ and $\rm I_{3}$ by making use of the following relation  (Eq. (3.11) of \citealt{Goldberg:1967})
\begin{eqnarray}
_{s}Y_{lm}(\hat q)=(-1)^{m}\sqrt{\frac{2l+1}{4\pi}}D_{-ms}^{l}(\hat q,\rho(\hat q))
\end{eqnarray}
for the spin-$s$ spherical harmonics. The integral $\rm I_{2}$ can be expressed as
\begin{eqnarray}
\rm I_{2}&=&(-1)^{m}\sqrt{\frac{2l+1}{4\pi}}\int d\Omega_{\hat q}D_{-m2}^{l}(\hat q,\rho(\hat q))D_{m'M}^{l'}(\hat q,\rho(\hat q)),
\end{eqnarray}
and from Eq. (1), Section 4.6 of \cite{Varshalovich:1988}, the expression reduces to the following form
\begin{eqnarray}
\rm I_{2}&=&(-1)^{m}\sqrt{\frac{2l+1}{4\pi}}\sum_{L=\vert l-l'\vert}^{l+l'}C_{l-m l' m'}^{L(-m+m')}C_{l2 l' M}^{L(2+M)}  \int d\Omega_{\hat q}D_{(-m+m')(2+M)}^{L}(\hat q,\rho(\hat q)).
\end{eqnarray}
We follow the same procedure to evaluate the integral $\rm I_{3}$ and find
\begin{eqnarray}
\rm I_{3}&=&(-1)^{m}\sqrt{\frac{2l+1}{4\pi}}\int d\Omega_{\hat q}D_{-m-2}^{l}(\hat q,\rho(\hat q))D_{m'M}^{l'}(\hat q,\rho(\hat q)),
\end{eqnarray}
which after expansion of the product of Wigner-D functions leads to
\begin{eqnarray}
\rm I_{3}&=&(-1)^{m}\sqrt{\frac{2l+1}{4\pi}}\sum_{L=\vert l-l'\vert}^{l+l'}C_{l-m l' m'}^{L(-m+m')}C_{l-2 l' M}^{L(-2+M)}  \int d\Omega_{\hat q}D_{(-m+m')(-2+M)}^{L}(\hat q,\rho(\hat q)).
\end{eqnarray}

\section{Decomposition of the bias matrix}
\label{appendix: Decomposition}
 From Eq. (\ref{eq: Allprime_fmprimeN}) we can decompose the general form of the bias matrix corresponding to the beam harmonic transform expansion ($m''=0, \pm2, \pm4 $ for the temperature and $M'=\pm2, \pm4, \pm6 $ for the E-mode) as follows
 \begin{eqnarray}
 A_{ll'}^{TE}(term\ 1)&=&\pi b_{l'0}^{T}b_{l'2}^{E}\sum_{m=-l}^{l} \sum_{L=\vert l-l'\vert}^{l+l'}C_{l-m l' m}^{L 0}C_{l 0 l' 0}^{L 0}\sum_{M=-L}^{L}d_{0M}^{L}(\frac{\pi}{2})d_{M0}^{L}(\frac{\pi}{2})f_{0M} \sum_{L'=\vert l-l'\vert}^{l+l'}C_{l-m l' m}^{L' 0}\nonumber \\
 &\times&\left. \left[ C_{l2 l' -2}^{L'0}\sum_{N=-L'}^{L'} d_{0N}^{L'}(\frac{\pi}{2})d_{N0}^{L'}(\frac{\pi}{2}) f_{0N}+ C_{l-2 l' -2}^{L'-4}\sum_{N=-L'}^{L'} d_{0N}^{L'}(\frac{\pi}{2})d_{N-4}^{L'}(\frac{\pi}{2}) f_{-4N}\right. \right.\nonumber\\
  &+&\left. C_{l2 l' 2}^{L'4}\sum_{N=-L'}^{L'} d_{0N}^{L'}(\frac{\pi}{2})d_{N4}^{L'}(\frac{\pi}{2}) f_{4N}+ C_{l-2 l' 2}^{L'0}\sum_{N=-L'}^{L'} d_{0N}^{L'}(\frac{\pi}{2})d_{N0}^{L'}(\frac{\pi}{2}) f_{0N}\right],
 \end{eqnarray}
 
  \begin{eqnarray}
 A_{ll'}^{TE}(term\ 2)&=&\pi b_{l'2}^{T}b_{l'2}^{E}\sum_{m=-l}^{l}\sum_{L=\vert l-l'\vert}^{l+l'}C_{l-m l' m}^{L 0}\left(  C_{l 0 l' -2}^{L -2}\sum_{M=-L}^{L}d_{0M}^{L}(\frac{\pi}{2})d_{M-2}^{L}(\frac{\pi}{2})f_{-2M}+ C_{l 0 l' 2}^{L 2}\sum_{M=-L}^{L}d_{0M}^{L}(\frac{\pi}{2})d_{M2}^{L}(\frac{\pi}{2})f_{2M}\right)\nonumber \\ 
 &\times &\left. \left. \sum_{L'=\vert l-l'\vert}^{l+l'}C_{l-m l' m}^{L' 0}\left[ C_{l2 l' -2}^{L'0}\sum_{N=-L'}^{L'} d_{0N}^{L'}(\frac{\pi}{2})d_{N0}^{L'}(\frac{\pi}{2}) f_{0N}+ C_{l-2 l' -2}^{L'-4}\sum_{N=-L'}^{L'} d_{0N}^{L'}(\frac{\pi}{2})d_{N-4}^{L'}(\frac{\pi}{2}) f_{-4N}\right. \right. \right. \nonumber\\
 &+&\left. C_{l2 l' 2}^{L'4}\sum_{N=-L'}^{L'} d_{0N}^{L'}(\frac{\pi}{2})d_{N4}^{L'}(\frac{\pi}{2}) f_{4N}+ C_{l-2 l' 2}^{L'0}\sum_{N=-L'}^{L'} d_{0N}^{L'}(\frac{\pi}{2})d_{N0}^{L'}(\frac{\pi}{2}) f_{0N}\right],
 \end{eqnarray}
 \begin{eqnarray}
 A_{ll'}^{TE}(term\ 3)&=&\pi b_{l'0}^{T}b_{l'4}^{E}\sum_{m=-l}^{l} \sum_{L=\vert l-l'\vert}^{l+l'}C_{l-m l' m}^{L 0}C_{l 0 l' 0}^{L 0}\sum_{M=-L}^{L}d_{0M}^{L}(\frac{\pi}{2})d_{M0}^{L}(\frac{\pi}{2})f_{0M} \sum_{L'=\vert l-l'\vert}^{l+l'}C_{l-m l' m}^{L' 0}\nonumber \\
 &\times &\left. \left[ C_{l2 l' -4}^{L'-2}\sum_{N=-L'}^{L'} d_{0N}^{L'}(\frac{\pi}{2})d_{N-2}^{L'}(\frac{\pi}{2}) f_{-2N}+ C_{l-2 l' -4}^{L'-6}\sum_{N=-L'}^{L'} d_{0N}^{L'}(\frac{\pi}{2})d_{N-6}^{L'}(\frac{\pi}{2}) f_{-6N}\right. \right.\nonumber\\
  &+&\left. C_{l2 l' 4}^{L'6}\sum_{N=-L'}^{L'} d_{0N}^{L'}(\frac{\pi}{2})d_{N6}^{L'}(\frac{\pi}{2}) f_{6N}+ C_{l-2 l' 4}^{L'2}\sum_{N=-L'}^{L'} d_{0N}^{L'}(\frac{\pi}{2})d_{N2}^{L'}(\frac{\pi}{2}) f_{2N}\right],
 \end{eqnarray}
 \begin{eqnarray}
 A_{ll'}^{TE}(term\ 4)&=&\pi b_{l'4}^{T}b_{l'2}^{E}\sum_{m=-l}^{l}\sum_{L=\vert l-l'\vert}^{l+l'}C_{l-m l' m}^{L 0}\left(  C_{l 0 l' -4}^{L -4}\sum_{M=-L}^{L}d_{0M}^{L}(\frac{\pi}{2})d_{M-4}^{L}(\frac{\pi}{2})f_{-4M}+ C_{l 0 l' 4}^{L 4}\sum_{M=-L}^{L}d_{0M}^{L}(\frac{\pi}{2})d_{M4}^{L}(\frac{\pi}{2})f_{4M}\right)\nonumber \\ 
 &\times &\left. \left. \sum_{L'=\vert l-l'\vert}^{l+l'}C_{l-m l' m}^{L' 0}\left[ C_{l2 l' -2}^{L'0}\sum_{N=-L'}^{L'} d_{0N}^{L'}(\frac{\pi}{2})d_{N0}^{L'}(\frac{\pi}{2}) f_{0N}+ C_{l-2 l' -2}^{L'-4}\sum_{N=-L'}^{L'} d_{0N}^{L'}(\frac{\pi}{2})d_{N-4}^{L'}(\frac{\pi}{2}) f_{-4N}\right. \right. \right. \nonumber\\
 &+&\left. C_{l2 l' 2}^{L'4}\sum_{N=-L'}^{L'} d_{0N}^{L'}(\frac{\pi}{2})d_{N4}^{L'}(\frac{\pi}{2}) f_{4N}+ C_{l-2 l' 2}^{L'0}\sum_{N=-L'}^{L'} d_{0N}^{L'}(\frac{\pi}{2})d_{N0}^{L'}(\frac{\pi}{2}) f_{0N}\right], 
 \end{eqnarray}
  \begin{eqnarray}
 A_{ll'}^{TE}(term\ 5)&=&\pi b_{l'0}^{T}b_{l'6}^{E}\sum_{m=-l}^{l} \sum_{L=\vert l-l'\vert}^{l+l'}C_{l-m l' m}^{L 0}C_{l 0 l' 0}^{L 0}\sum_{M=-L}^{L}d_{0M}^{L}(\frac{\pi}{2})d_{M0}^{L}(\frac{\pi}{2})f_{0M} \sum_{L'=\vert l-l'\vert}^{l+l'}C_{l-m l' m}^{L' 0}\nonumber \\
 &\times &\left. \left[ C_{l2 l' -6}^{L'-4}\sum_{N=-L'}^{L'} d_{0N}^{L'}(\frac{\pi}{2})d_{N-4}^{L'}(\frac{\pi}{2}) f_{-4N}+ C_{l-2 l' -6}^{L'-8}\sum_{N=-L'}^{L'} d_{0N}^{L'}(\frac{\pi}{2})d_{N-8}^{L'}(\frac{\pi}{2}) f_{-8N}\right. \right.\nonumber\\
  &+&\left. C_{l2 l' 6}^{L'8}\sum_{N=-L'}^{L'} d_{0N}^{L'}(\frac{\pi}{2})d_{N8}^{L'}(\frac{\pi}{2}) f_{8N}+ C_{l-2 l' 6}^{L'4}\sum_{N=-L'}^{L'} d_{0N}^{L'}(\frac{\pi}{2})d_{N4}^{L'}(\frac{\pi}{2}) f_{4N}\right].
 \end{eqnarray}
\end{document}